\documentclass[pra,aps,twocolumn,10pt,superscriptaddress]{revtex4-2}
\usepackage{amsmath,amsfonts,amssymb,graphics,graphicx,epsfig,color,times}

\usepackage{graphicx}
\usepackage{dcolumn}
\usepackage{bm}
\usepackage{hyperref}
 \usepackage{xcolor}
 \usepackage{lineno}

\begin{document}

\preprint{APS/123-QED}

\title{\bf Accelerating multipartite entanglement generation in non-Hermitian superconducting qubits}

\author{Chimdessa Gashu Feyisa}
\affiliation{Institute of Atomic and Molecular Sciences, Academia Sinica, Taipei 10617, Taiwan}
\affiliation{Molecular Science and Technology Program, Taiwan International Graduate Program, Academia Sinica, Taiwan}
\affiliation{Department of Physics, National Central University, Taoyuan 320317, Taiwan}
\author{J. S. You}
\affiliation{Department of Physics, National Taiwan Normal University, Taipei 11677, Taiwan}
\author{Huan-Yu Ku}
\email{huan.yu@ntnu.edu.tw}
\affiliation{Department of Physics, National Taiwan Normal University, Taipei 11677, Taiwan}
\author{H. H. Jen}
\email{sappyjen@gmail.com}
\affiliation{Institute of Atomic and Molecular Sciences, Academia Sinica, Taipei 10617, Taiwan}
\affiliation{Molecular Science and Technology Program, Taiwan International Graduate Program, Academia Sinica, Taiwan}
\affiliation{Physics Division, National Center for Theoretical Sciences, Taipei 10617, Taiwan}
\date{\today}

\begin{abstract}
Open quantum systems are susceptible to losses in information, energy, and particles due to their surrounding environment. One novel strategy to mitigate these losses is to transform them into advantages for quantum technologies through tailored non-Hermitian quantum systems. In this work, we theoretically propose a fast generation of multipartite entanglement in non-Hermitian qubits. Our findings reveal that weakly coupled non-Hermitian qubits can accelerate multiparty entanglement generation by thousands of times compared to Hermitian qubits, in particular when approaching the $2^n$-th order exceptional points of $n$ qubits in the ${\cal P}{\cal T}-$ symmetric regime. Furthermore, we show that Hermitian qubits can generate GHZ states with a high fidelity more than $0.9995$ in a timescale comparable to that of non-Hermitian qubits, but at the expense of intense driving and large coupling constant. Our approach is scalable to a large number of qubits, presenting a promising pathway for advancing quantum technologies through the non-Hermiticity and higher-order exceptional points in many-body quantum systems.
\end{abstract}

\maketitle


\section*{\label{I} Introduction}
Quantum physics in the Hermitian realm provides a genuine description of the dynamics of the closed quantum systems. Hermiticity ensures the conservation of essential quantities such as probability, energy, particle number, and information \cite{ar00}. However, when quantum systems interact with their surroundings, one frequently employs the Lindblad master equation \cite{ar01}. It includes a Hermitian term that drives the unitary and coherent evolution of the system, along with additional non-Hermitian terms accounting for dissipations.

A broader framework for dissipative quantum systems has been developed using non-Hermitian Hamiltonians \cite{ar02, ar24, ar23, ar1, ar2, ar3}. These Hamiltonians exhibit intriguing physical phenomena such as imaginary eigenvalues, biorthogonal states, and exceptional points (EPs) \cite{ar21, ar515, ar25,Lin2024arxiv}. EPs are complex branching singularities in a parameter space where both eigenvalues and eigenvectors coalesce. These singular points are unique to non-Hermitian systems and lead to heightened sensitivity to applied perturbations, causing dramatic changes in the system's behavior \cite{ar5, ar03}. This property has spurred significant interest in developing quasi-classical systems capable of exhibiting EPs, with applications in ultra-sensitive sensing \cite{ar501, ar502, ar503}, wave transport management \cite{ar504, ar505}, and single-mode lasing operations \cite{ar506, ar507}. EPs have also been experimentally demonstrated in various quantum systems, including superconducting qubits \cite{ar2, ar3, ar301}, trapped ions \cite{ar509, ar510}, thermal atom ensembles \cite{ar511}, and cold atoms \cite{ar512}, offering tremendous quantum advantages \cite{ar513, ar514, ar515, ar516, ar517, ar518, ar519,JuPRA2019}.

Among these benefits, the dynamics of entanglement around EPs have recently attracted much attention \cite{ar508, ar04, ar1, ar6, ar26, ar41}. For example, second-order EP has been identified as the location of Bell state generation between non-Hermitian qubit and Hermitian qubits \cite{ar6}. In addition, fourth-order EPs in two non-Hermitian qubits have been shown to accelerate bipartite entanglement generation \cite{ar1}. Furthermore, the trade-off relationship between the degree of entanglement and the success rate has recently been investigated in non-Hermitian qubits with (un)balanced gain and loss \cite{ar26}. Despite these pioneering studies on the interplay between bipartite entanglement and EPs, very little is known about the role of higher-order EPs in multipartite entanglement. 

Multipartite entanglement has been implemented in various protocols involving Hermitian systems and gate operations \cite{ar520,ar521,ar522,ar523,ar526,ar36,ar37}. In general, it provides quantum advantages over bipartite entanglement \cite{ar524,ar525} e.g., in quantum networks \cite{ar29, ar32}, quantum secret sharing \cite{ar30,ar31,ar32}, quantum key distributions \cite{ar33,ar34}, and quantum thermodynamic tasks \cite{ar35}. By arranging non-Hermitian and Hermitian qubits in desired configurations, multipartite entanglement allows for the adjustment of higher-order EPs. In principle, higher-order EPs could further amplify the effects of external perturbations and initial conditions. Therefore, it is crucial to understand the advantages of higher-order EPs in multipartite entanglement over (1) conventional Hermitian counterparts and (2) gate operations that require numerous and time-consuming SWAP operations.

In this work, we investigate multipartite entanglement in driven, dissipative, and weakly coupled non-Hermitian transmon qubits, which have been realized experimentally in Refs. \cite{ar2,ar3}. Without inter-qubit coupling, the $2^n$-th order EPs emerge among $n$ qubits due to the rivalry between dissipation-induced non-Hermiticity and resonant driving fields. We make quantitative analysis of entanglement dynamics near these higher-order EPs and in the Hermitian limit using entanglement entropy \cite{ar7} and residual three-tangle \cite{ar10, ar13}, alongside the dynamics of probability amplitudes and phases to understand the physics of multipartite entanglement generation. Our result shows that the simultaneous rearrangement of populations and phases in three non-Hermitian qubits quickly builds robust tripartite entanglement near the eighth-order EP. More intriguingly, four non-Hermitian qubits can be entangled even more rapidly due to the influence of higher-order EPs involved in them. Conversely, Hermitian qubits cost strong driving and relatively large inter-qubit coupling to develop high-fidelity multipartite entangled states within a comparable timescale of the non-Hermitian qubits. 
\section*{\label{II} Results}
\subsection*{\label{II0} Emergence of higher-order EPs}
\begin{figure}
	\begin{center}
		\includegraphics[width=.5\textwidth]{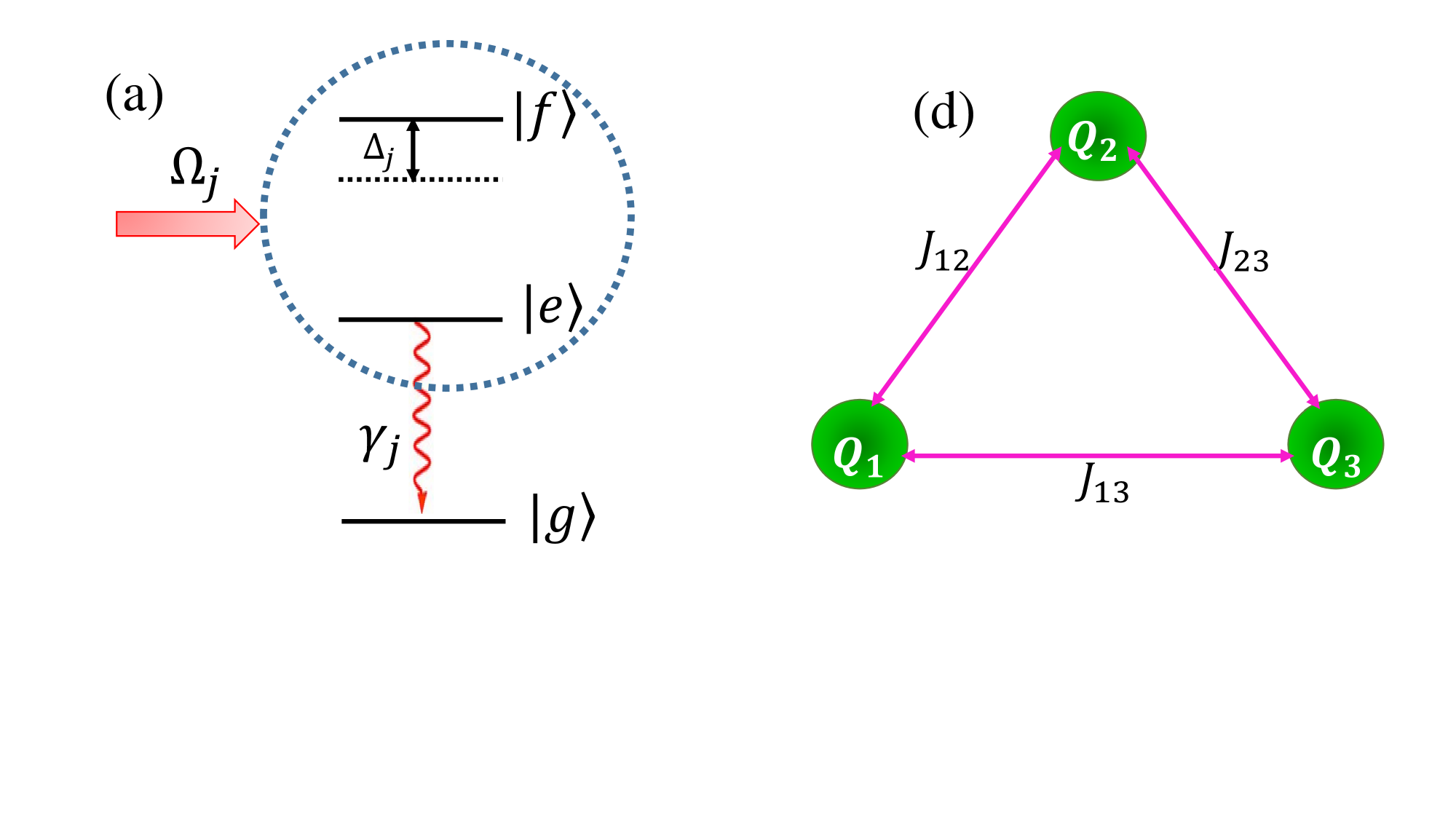}\\\vspace{-1.65cm}
		\includegraphics[width=.5\textwidth]{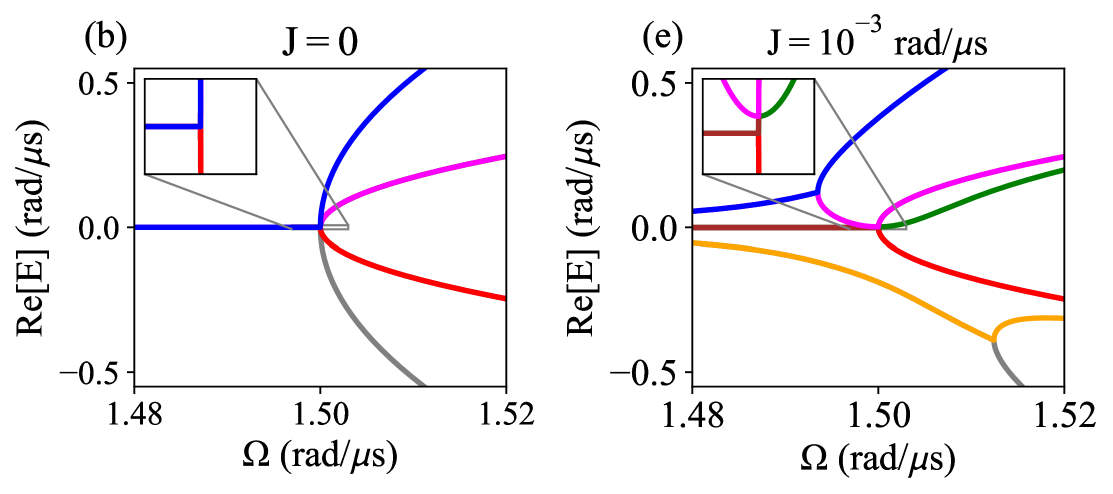}\\
		\hspace{-2.75cm}{(c)\hspace{3.75cm}(f)}\\
		\includegraphics[width=.5\textwidth]{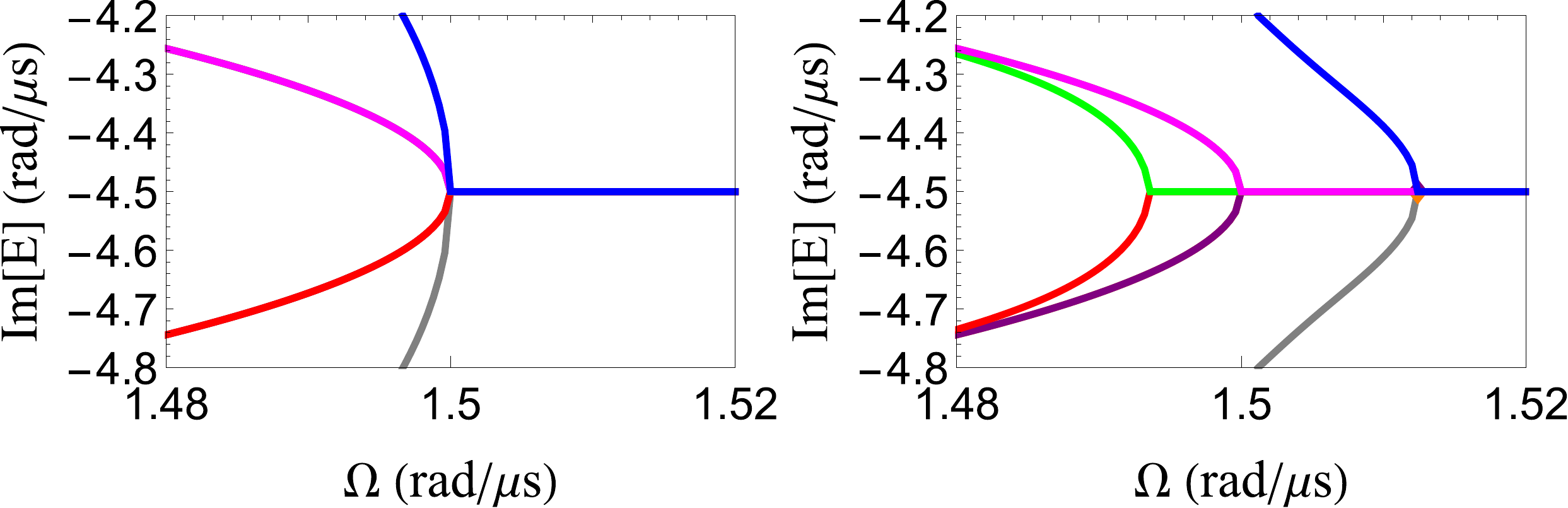}
		\caption{Schematics of a single qubit level structure, three coupled non-Hermitian qubits, and associated high-order EPs. (a) A single non-Hermitian qubit is constructed within the manifold $\{|e\rangle, |f\rangle\}$ of a three-level system by selectively removing the ground state $|g\rangle$ through driven and dissipative processes. The parameter $\Omega_j$ represents the amplitude of the laser drive coupling transitions within the qubit manifold and is detuned by frequency $\Delta_j$, while $\gamma_j$ denotes the decay rate of the state $|e\rangle_j$. Here, we assume that the three qubits share identical driving amplitudes $\Omega_j=\Omega$, decay rates $\gamma_j=\gamma=6$ rad/$\mu$s, and inter-qubit coupling $J_{jk}=J$. (b) Three coupled non-Hermitian qubits (green full circles) with cyclic interactions. The real and imaginary components of the eigenvalues of the qubits are respectively illustrated in (c) and (d) for $J=0$ and (e) and (f) for $J=10^{-3}$ rad/$\mu$s. For $J=0$, the three-qubit system driven with $\Omega=1.5$ rad/$\mu$s exhibits an eighth-order EP, which reduces to third-order and fourth-order EPs for $J=10^{-3}$ rad/$\mu$s. These EPs are further illustrated in the magnified insets of panels (b) and (e).}\label{f0}
	\end{center}
\end{figure}

We consider a transmon circuit consisting of a capacitor and an inductor connected by a superconducting wire \cite{ar1, ar2, ar3, ar4}. The key component of this circuit is the Josephson Junction, which provides non-linear inductance, imparting an anharmonic character to the transmon circuit \cite{ar2}, in contrast to the harmonic oscillation of an ideal LC circuit \cite{ar4}. Controlling the circuit elements allows us to access three energy levels of the transmon circuit successively named ground state $|g\rangle$, first excited state $|e\rangle$, and second excited state $|f\rangle$. In fact, the circuit resembles a qutrit system depicted in Fig. \ref{f0}(a). The distinct energy spacings of the circuit facilitate the realization of a non-Hermitian qubit in the subspace $\{|e\rangle, |f\rangle\}$ via driven-dissipative processes, while the ground state $|g\rangle$ serves as an environment external to the qubit. To achieve the qubit's functionality in the chosen subspace, it is necessary to ensure faster decay of the state $|e\rangle$ compared to the state $|f\rangle$, and this hierarchy of dissipation rates can be monitored by an impedance-mismatching element, which amplifies or suppresses electromagnetic radiation mode in the three-dimensional microwave cavity \cite{ar2, ar3}. 

These systems provide insights into entanglement dynamics within the framework of non-Hermitian Hamiltonians \cite{ar1}. These Hamiltonians are effective for modeling realistic scenarios involving dissipative quantum systems. In this context, we consider weakly coupled non-Hermitian qubits described by the total non-Hermitian Hamiltonian
\begin{eqnarray}
\hat{H}&=&\sum^{n}_{j=1}\bigg[(\Delta_j-\frac{i\gamma_j}{2})\hat{\sigma}_{j}\hat{\sigma}^{\dagger}_{j}+\Omega_{j}\hat{\sigma}^{x}_{j}\bigg]\nonumber\\&+&\sum^{n}_{j\neq k}\sum^{n}_{k=1}J_{jk}\big(\hat{\sigma}^{\dagger}_{j}\hat{\sigma}_{k}+\hat{\sigma}_{j}\hat{\sigma}^{\dagger}_{k}\big),\label{e1}\end{eqnarray}\label{m0}where the terms in the square bracket represent the individual non-Hermitian Hamiltonian for each qubit, while $\Omega_j$ denotes the driving amplitude, $\Delta_j$ represents the detuning frequency, and $\gamma_j$ is the dissipation rate of the level $|e\rangle_j$. The Pauli matrices for the qubit are given by $\hat{\sigma}_j^x=\hat{\sigma}_j^\dagger + \hat{\sigma}_j$, with $\hat{\sigma}_j^\dagger=|f\rangle_j \langle e|$ and $\hat{\sigma}_j=|e\rangle_j \langle f|$. The second term in Eq.~(1) describes the inter-qubit coupling with constant $J_{jk}$. For $\Delta_j=0$, our model obeys the passive ${\cal P}{\cal T}-$symmetry i.e., $[{\cal P}{\cal T}, \hat H_{{\cal P}{\cal T}}]=0,$~\cite{ar1, ar40} where $\hat H_{{\cal P}{\cal T}}=\hat H(\Delta_j=0)+\sum^{n}_{j=1}(i\gamma_j)/4$. Here, ${\cal P}$ and ${\cal T}$ indicate the parity and the time-reversal operators, respectively. 

Before showing our main result (i.e., the fast generation of multipartite entanglement in $n$ qubits), we first discuss their EPs, and then we address advantages of these EPs in multipartite entanglement generation later. A single non-Hermitian qubit in Fig. \ref{f0}(a) exhibits a second-order EP at $\Omega_j=\gamma_j/4$ \cite{ar2}, and two identical non-Hermitian qubits exhibit fourth order EP at the same Rabi frequency \cite{ar1}. In addition, Figs. \ref{f0}(b) and \ref{f0}(c) show the real and imaginary parts of the eigenvalues for three identical, uncoupled non-Hermitian qubits with $\Delta_j=0$. At $\Omega=\gamma/4=1.5$ rad/$\mu$s (referred to as $\Omega_{\rm EP}$ hereafter), the qubits exhibit an eighth-order EP, where all eigenvalues converge to $E_{\rm EP}=-3i\gamma/4$, and all eigenvectors coalesce into $|\varphi_{\rm EP}\rangle$ (see Supplementary Note A for more detail). This result generalizes to $(2^n)^{\rm th}$ order EPs for $n$ uncoupled non-Hermitian qubits, with the EP order matching the Hilbert space dimensions. These higher-order EPs arise from the rivalry between dissipation-induced non-Hermiticity and resonant driving field effects. In practice, they can be tuned by varying the resonant Rabi frequency while keeping the non-Hermiticity parameter \cite{ar2}.

Figure \ref{f0}(d) illustrates three identical non-Hermitian qubits weakly coupled to each other with a uniform coupling constant $J_{jk}=J$. This coupling acts as an external perturbation to the system and reduces the order of EPs at $\Omega_{\rm EP}$ due to degeneracy lifting. For example, with a coupling strength of $J = 10^{-3}$ rad/$\mu$s, an eighth-order EP is reduced to fourth- and third-order EPs at $\Omega_{\rm EP}$. Furthermore, two second-order EPs emerge on either side of $\Omega_{\rm EP}$ as shown in Figs. \ref{f0}(e) and \ref{f0}(f). These second-order EPs persist even in the strong coupling regime ($J>\gamma$), although increasing the coupling strength completely lifts the degeneracy at $\Omega_{\rm EP}$. In contrast, lower coupling strengths such as $J=10^{-4}$, $10^{-5}$, and $10^{-6}$ rad/$\mu$s result in a fifth-order EP at $\Omega_{\rm EP}$, with the response of the system weakening as the coupling strength decreases. Around the higher-order EPs, the response of energy eigenvalues can be enhanced by several orders of magnitude \cite{ar39, ar5}, resulting in the heightened sensitivity of the non-Hermitian qubits to external perturbations. Later, we will show that this effect can indeed boost multipartite entanglement generation.
\subsection*{\label{III} EP-induced multipartite entanglement generation}
\begin{figure}
	\centering
	\hspace{-0.5cm}(a)\hspace{4cm}\textcolor{white}{(b)} 
	\includegraphics[width=0.37\textwidth]{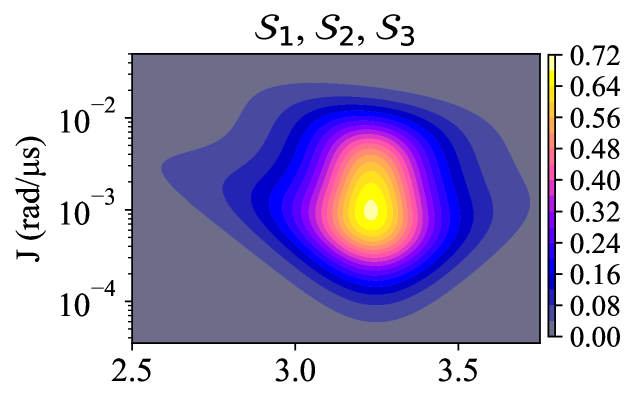}\\ 
	\hspace{-0.5cm}(b)\hspace{4cm}\textcolor{white}{(b)} 
	\includegraphics[width=0.37\textwidth]{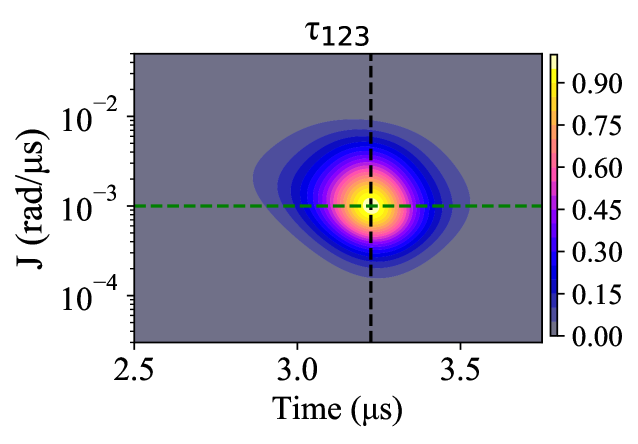} 
	\caption{Tripartite entanglement dynamics of three coupled non-Hermitian qubits from $|\psi(0)\rangle=2^{-3/2}(|f\rangle-i|e\rangle)^{\otimes3}$. The density plots depict evolutions of (a) entanglement entropies ${\cal S}_{j}$ of reduced qubits and (b) three-tangle $\tau_{123}$ (see Methods) as functions of time and interaction strength $J$ for $\Omega=1.576$ rad/$\mu$s and $\gamma=6$ rad/$\mu$s. The color scales in each plot illustrate the magnitudes of the respective quantifiers. The crossing point of green-dashed and black-dashed lines in the three-tangle marks EP-induced optimal entanglement at $t\approx3.233$ $\mu$s and $J\approx10^{-3}$ rad/$\mu$s.}\label{f1}
\end{figure} 
\begin{figure*}[!t]
	\centering
	\hspace{-3.5cm}(a)\hspace{4cm}(b)\hspace{5cm}(c)
	\includegraphics[width=0.87\textwidth]{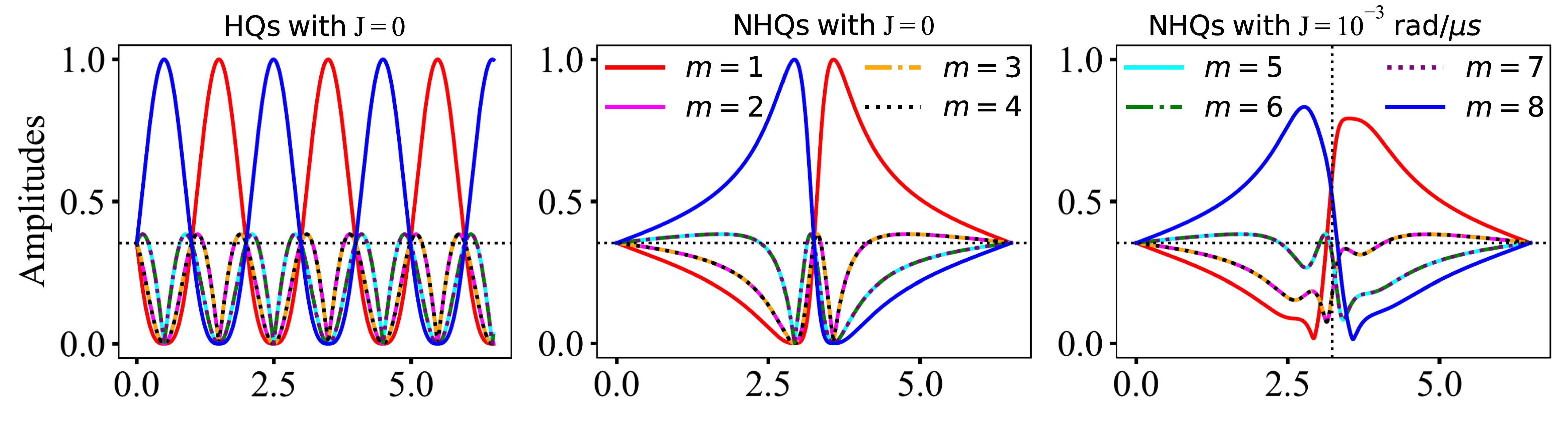}\\
	\includegraphics[width=0.87\textwidth]{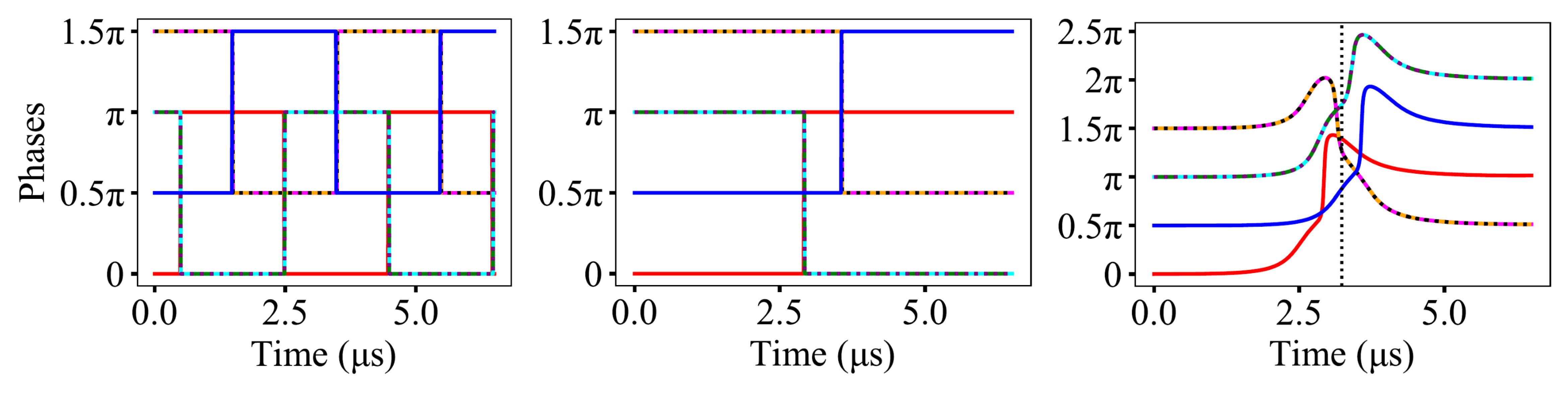}\\
	\hspace{-2cm}(d)\hspace{3.75cm}(e)\hspace{3.5cm}(f)\hspace{3.75cm}(g)
	\includegraphics[width=0.87\textwidth]{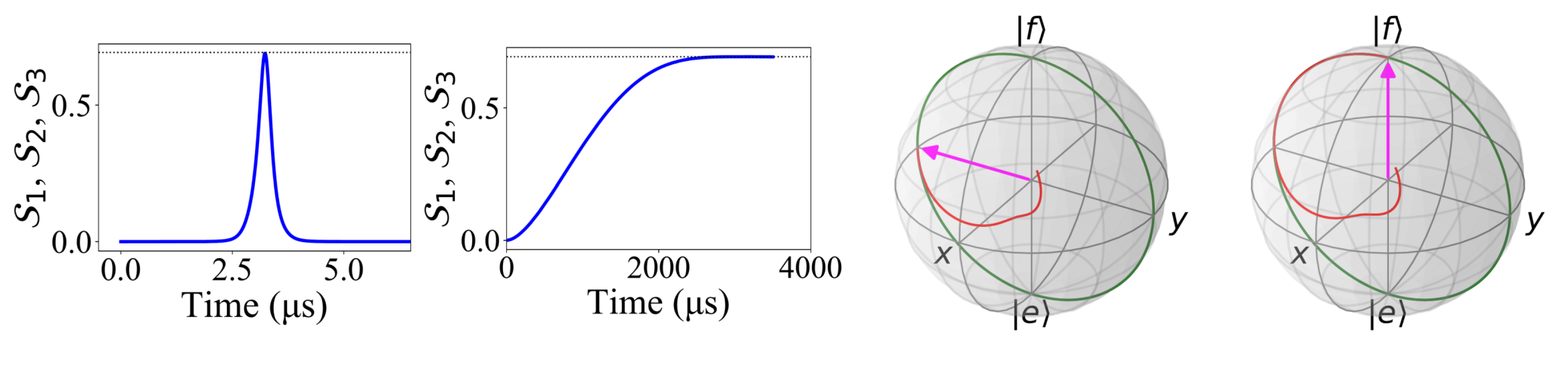}
	\caption{Dynamics of Hermitian and non-Hermitian qubits. (a)-(c) Time evolutions of probability amplitudes $|\alpha_{m}|$ and phases Arg$(\alpha_{m})$ for three Hermitian qubits (HQs) for $J=0$ (a) and non-Hermitian qubits (NHQs) for $J=0$ (b) and $J=10^{-3}$ rad/$\mu$s (c). Probability amplitudes and phases are shown for $m=1$ (solid red), $2$ (solid magenta), $3$ (dash-dotted orange), $4$ (dotted black), $5$ (solid cyan), $6$ (dash-dotted green), $7$ (dotted purple), and $8$ (solid blue). These correspond to the three-qubit state $|\psi(t)\rangle = \alpha_{1}(t)|fff\rangle + \alpha_{2}(t)|ffe\rangle + \alpha_{3}(t)|fef\rangle + \alpha_{4}(t)|eff\rangle + \alpha_{5}(t)|fee\rangle + \alpha_{6}(t)|efe\rangle + \alpha_{7}(t)|eef\rangle + \alpha_{8}(t)|eee\rangle$ constructed from the coherent state $|\psi(0)\rangle = 2^{-3/2}(|f\rangle - i|e\rangle)^{\otimes 3}$. Horizontal dotted lines in the upper figures represent equal probability amplitudes ($|\alpha_{m}|=1/(2\sqrt{2})$) for the three-qubit state. The computational bases of identical qubits with the same number of excitations are degenerate: $|fee\rangle$, $|efe\rangle$, and $|eef\rangle$ form one set of three degenerate bases, while $|ffe\rangle$, $|fef\rangle$, and $|eff\rangle$ form another set of three degenerate bases (see text). (d) and (e) Entanglement entropies of the reduced non-Hermitian qubits (d) and Hermitian qubits (e) for $J=10^{-3}$ rad/$\mu$s, with the horizontal dotted lines representing the highest value ${\cal S}_j=\log{2}$ for GHZ states (see Methods). (f) and (g) Evolution of a reduced qubit on the Bloch sphere: from $|\psi(0)\rangle=2^{-3/2}(|f\rangle-i|e\rangle)^{\otimes 3}$ (f) and $|\psi(0)\rangle = |fff\rangle$ (g). Green trajectories show the evolution of the reduced Hermitian qubit with $J=0$ and $J=10^{-3}$ rad/$\mu$s, and the reduced non-Hermitian qubit without coupling for $t=6.5$ $\mu$s, while red trajectories indicate the non-Hermitian qubit with $J=10^{-3}$ rad/$\mu$s for $t=3.232$ $\mu$s (f) and $t=5.325$ $\mu$s (g). The magenta arrows show the initial states of the qubits. Other parameters are the same as in Fig. \ref{f1}.}\label{f2}
\end{figure*}

We focus on the dynamics of non-Hermitian qubits in the ${\cal P}{\cal T}$-symmetry preserving regime \cite{ar5,ar14,ar40}, where $\Omega\geq\Omega_{\rm EP}$. In this regime, the qubit system exhibits equal imaginary eigenvalues, which are removed from the state $|\psi(t)\rangle=e^{-i\hat{H}t/\hbar}|\psi(0)\rangle/\sqrt{\langle\psi(0)|e^{i\hat{H^\dag}t/\hbar}e^{-i\hat{H}t/\hbar}|\psi(0)\rangle}$ of the composite system through normalization conditions. Meanwhile, the distinct real eigenvalues significantly influence the system dynamics (see Figs. \ref{f0}(d) and \ref{f0}(b)). We can then write the normalized state of $n$ qubits as $|\psi(t)\rangle=\sum^{2^n}_{m=1}\alpha_m(t)|\delta_m\rangle$, where $\alpha_m(t)$ signifies the complex probability amplitude, and $|\delta_m\rangle$ is the computational basis (see Methods for further details).

We aim to generate multipartite GHZ states \cite{ar38}, which can be distinguished from any other states using the entanglement entropy \cite{ar7} ${\cal S}_{j}=-{\rm Tr}[\hat\rho_j(t)\log{\hat\rho_j(t)}]$, where $\hat{\rho}_{j}(t)$ stands for the reduced state of qubit $j$ for $j=1$, $2$, ..., $n$ (see Methods). Each qubit subsystem in GHZ states achieves the maximum entanglement entropy ${\cal S}_j=\log{2}$. This can be further confirmed by the genuine residual three-tangle $\tau_{123}$ for tripartite system when $\tau_{123}=1$ \cite{ar10,ar13}. 

We first investigate the entanglement developed in three-qubit system in the ${\cal P}{\cal T}$-symmetric regime $\Omega\geq\Omega_{\rm EP}$. In Figs. \ref{f1}(a) and \ref{f1}(b), we illustrate the dynamics of entanglement entropy and residual three-tangle as functions of time and inter-qubit coupling for a specific driving amplitude $\Omega=1.576$ rad/$\mu$s. The quantifiers consistently indicate the generation of an optimal GHZ class from the initial coherent state $|\psi(0)\rangle=2^{-3/2}(|f\rangle-i|e\rangle)^{\otimes3}$ at $t\approx3.233$ $\mu$s and $J\approx10^{-3}$ rad/$\mu$s. At this point, the entanglement entropy of each reduced qubit is ${\cal S}_{j}\approx0.690$, and the three-tangle is ${\tau}_{123}\approx0.980$. This entangled state is highlighted by green-dashed and black-dashed lines intersecting at the central, brighter parts of the three-tangle (see Fig. \ref{f1}(b)). This indicates that for a specific driving amplitude and coupling constant, there exists a scenario where entanglement is significantly shared among all three qubits, with negligible bipartite entanglement between any two qubits (see Supplementary note B). This is a characteristic property of GHZ states \cite{ar10, ar13}. After the optimal entanglement generation at $t \approx 3.233$ $\mu$s in Fig. 2, the qubits evolve into product states. To understand the physics behind the GHZ state generation at $t\approx3.233$ $\mu$s and $J\approx10^{-3}$ rad/$\mu$s, we compare the dynamics of probability amplitudes $|\alpha_m(t)|$ and phases ${\rm Arg}[\alpha_m(t)]$ of the Hermitian and non-Hermitian qubits in Figs. \ref{f2}(a)-\ref{f2}(c) for $J=0$ and $J=10^{-3}$ rad/$\mu$s within a time span of $6.5$ $\mu$s (one period of evolution for non-Hermitian qubits). Hermitian qubits refer to a regime with $\gamma=0$ in the non-Hermitian Hamiltonian.

Figure \ref{f2}(a) illustrates that the amplitudes of the bases $|fff\rangle$ and $|eee\rangle$ of non-interacting Hermitian qubits oscillate periodically within $\pi/\Omega\approx 2$ $\mu$s. These oscillations result from coherent energy transfer between the qubits and the driving field, similar to Rabi oscillations of two-level atoms interacting with a quantized radiation mode \cite{ar12}. This coherent energy exchange results in a periodic evolution of the Hermitian qubits via the product states $|\psi(0)\rangle \rightarrow |eee\rangle \rightarrow |\psi^\ast(0)\rangle \rightarrow |fff\rangle \rightarrow |\psi(0)\rangle$ every $2$ $\mu$s.

Since local operation with classical communications is an allowed operation under the framework of the resource theory of entanglement~\cite{ar42, ar43, ar44}, we now discuss whether changing the local basis influences the generation of entangled states.
As the qubits jump from one basis state to another on the Bloch sphere, the state $|fff\rangle$ exchanges a discrete $\pi$-phase with the single-excitation degenerate bases $|fee\rangle$, $|efe\rangle$, and $|eef\rangle$ at $\pi/4\Omega\approx0.5$ $\mu$s, while the state $|eee\rangle$ exchanges an equal $\pi$-phase with the double-excitation degenerate bases $|ffe\rangle$, $|fef\rangle$, and $|eff\rangle$ at $3\pi/4\Omega\approx1.5$ $\mu$s. 
Furthermore, applying a small perturbation of the order $J=10^{-3}$ rad/$\mu$s does not substantially affect probability amplitudes of the Hermitian qubits within such a short time dynamics. However, it may slightly distort phase evolutions, which is insufficient for generating tripartite entanglement. This is because generating tripartite entanglement requires a simultaneous rearrangement of both populations and phases as will be clarified below.

We now demonstrate the dynamics of probability amplitudes and phases of the non-Hermitian qubits in Figs. \ref{f2}(b) and \ref{f2}(c) for $J=0$ and $J=10^{-3}$ rad$/\mu$s, respectively. Without coupling (see Fig. \ref{f2}(b)), the qubits evolve in a manner similar to Hermitian qubits, but the introduced non-Hermiticity distorts the oscillations and increases the period to $4\pi/\sqrt{16\Omega^2 - \gamma^2}\approx6.5$ $\mu$s compared to Hermitian qubits. In Fig. \ref{f2}(c) at a finite coupling $J=10^{-3}$ rad$/\mu$s, the dynamics of qubits at $t\approx3.233$ $\mu$s is significantly modified, where the amplitudes for $|fff\rangle$ and $|eee\rangle$ are increased, while the ones for the degenerate bases are decreased. This further triggers continuous phase evolutions instead of the discrete $\pi$-phase jumps seen without coupling. 

The rearrangement of both populations and phases highlights the fact that entanglement generation in three-qubit is fundamentally different from two-qubit cases, where all maximally entangled bipartite states are equivalent up to local change of bases \cite{ar13}. The vertical dashed lines in Fig. \ref{f2}(c) at $t\approx3.233$ $\mu$s indicate an entangled state generated at the peaks of entanglement entropies shown in Fig. \ref{f2}(d) (at the central bright regions in Fig. \ref{f1}). In contrast, achieving tripartite entanglement between three Hermitian qubits under the same parameter regimes requires several thousand microseconds, as depicted in Fig. \ref{f2}(e).
\begin{figure}
	\begin{center}
		\hspace{-1.75cm}(a)\hspace{4cm}\textcolor{white}{(b)}\\ 
		\includegraphics[width=.5\textwidth]{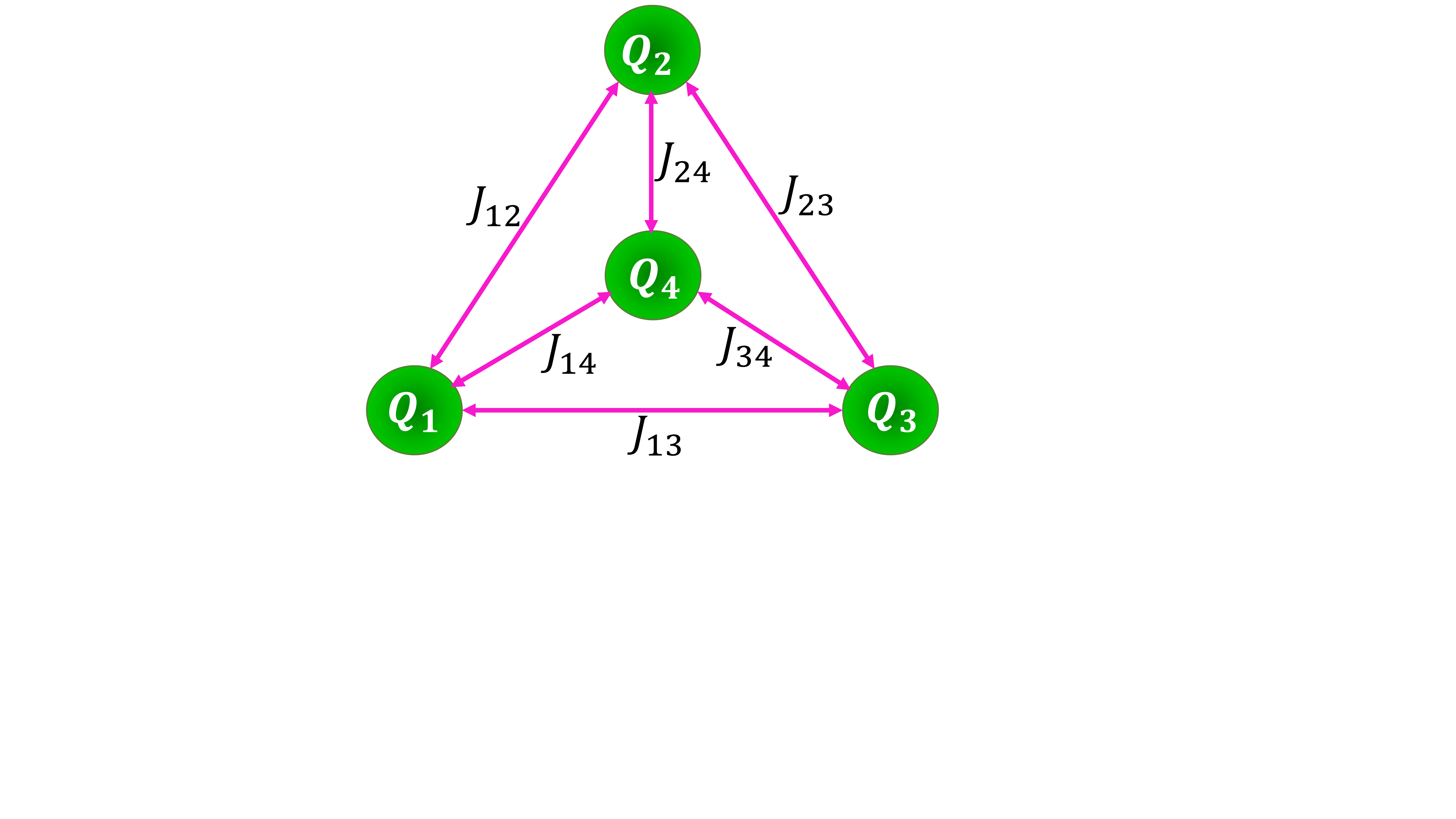}\\ \vspace{-2cm}
		\hspace{-2.75cm}(b)\hspace{4cm}\textcolor{white}{(b)}\\
		\includegraphics[width=.47\textwidth]{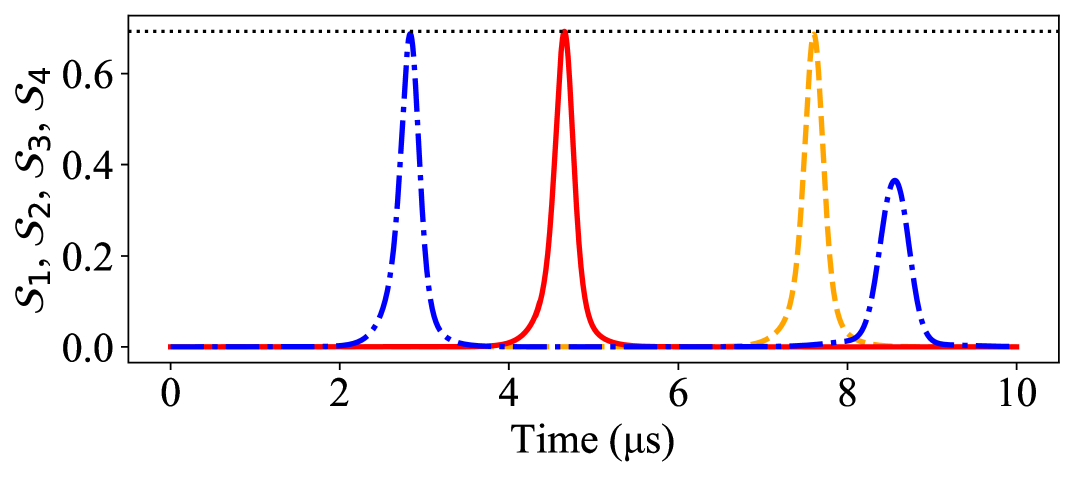}\\
		\caption{Entanglement generation in four coupled non-Hermitian qubits from the initial state $|\psi(0)\rangle=2^{-4/2}(|f\rangle-i|e\rangle)^{\otimes4}$. (a) Configurations of the qubits with all-to-all coupling. (b) Entanglement entropies ${\cal S}_j$ of the reduced non-Hermitian qubits for $\Delta=0$, $\{\Omega,J\}=\{1.514$ rad/$\mu$s,$10^{-5}$ rad/$\mu$s$\}$ (dashed orange), $\{\Omega,J\}=\{1.537$ rad/$\mu$s,$10^{-4}$ rad/$\mu$s$\}$ (solid red), and $\{\Omega,J\}=\{1.598$ rad/$\mu$s,$10^{-3}$ rad/$\mu$s$\}$ (dash-dotted blue). These different sets of coupling constants and resonant driving amplitudes are chosen at a fixed non-Hermiticity parameter $\gamma=6$ rad/$\mu$s and with respect to the location of higher-order EP at $\Omega_{\rm EP}$. They produce nearly maximal maximal entanglement entropies, highlighting four-qubit entanglement at different times.}\label{f4}
	\end{center}
\end{figure}

To gain further insights into the dynamics of the qubits, we visualize the evolution of reduced Hermitian and non-Hermitian qubits starting from the coherent state $|\psi(0)\rangle=2^{-3/2}(|f\rangle-i|e\rangle)^{\otimes3}$ (Fig. \ref{f2}(f)) and the product state $|\psi(0)\rangle=|fff\rangle$ (Fig. \ref{f2}(g)). The green trajectories in both figures represent the reduced Hermitian qubit with and without coupling, as well as the reduced non-Hermitian qubit without coupling. In these scenarios, both reduced qubits independently evolve on the surface of the Bloch sphere and remain in pure product states, indicating the absence of tripartite entanglement. For the Hermitian qubit with $J=10^{-3}$ rad$/\mu$s, the green trajectory suggests a regime where tripartite entanglement generation is not accessible due to the short qubit dynamics ($t=6.5$ $\mu$s) (see Fig. \ref{f2}(e)).

On the other hand, the non-Hermitian qubit with $J=10^{-3}$ rad$/\mu s$ evolves into a mixed state, suggesting significant tripartite entanglement generation. The corresponding purity values are $P_j\approx0.5033$ at $t\approx3.232$ $\mu$s for an initial coherent state and $P_j\approx0.512$ at $t\approx5.325$ $\mu$s for an initial state $|fff\rangle$ (see red trajectories in Figs. \ref{f2}(f) and \ref{f2}(g)). The lowest purity value, $P_1=P_2=P_3=0.5$, corresponds to maximally mixed states and indicates the standard three-party GHZ state. Notably, the initial coherent state $|\psi(0)\rangle=2^{-3/2}(|f\rangle-i|e\rangle)^{\otimes3}$ not only generates a robust genuine and bi-separable tripartite entangled state but also saves time and driving energy compared to the state $|\psi(0)\rangle=|fff\rangle$ (supplementary note B and C). This advantage arises due to the sensitivity of non-Hermitian qubits to initial conditions as well as applied perturbations, which is significantly beneficial for tasks requiring multiple qubits with higher-order EPs. 

Our system is scalable to $n$-qubit configurations, exemplified by a setup of four coupled non-Hermitian qubits shown in Fig. \ref{f4}(a). The entanglement entropies of each reduced-qubit system, depicted in Fig. \ref{f4}(b), can detect four-qubit GHZ-class states (Methods). For a given resonant driving amplitude $\Omega$, we can find an optimal value of coupling constant $J$ that generates four-qubit entanglement (see peaks of ${\cal S}_j$ in Fig. \ref{f4}(b)). The timing differences in entanglement generation are owing to the system's response to the combined effects of higher-order EPs, resonant driving, and inter-qubit coupling.

As the system approaches higher-order EPs located at $\Omega_{\rm EP}$, a weaker coupling strength is required to generate an optimal four-qubit entanglement. This is evident from the delayed peak of Fig. \ref{f4}(b). Increasing the coupling constant leads to faster entanglement generation slightly away from higher-order EPs. This is illustrated by the faster generation peaks. In addition, larger coupling constants enhance the robustness of the entangled states against off-resonant drivings (Supplementary note D). Thus, non-Hermitian qubits enable a design of specific driving protocols and inter-qubit couplings, allowing for the generation of multipartite entanglement unattainable with Hermitian qubits.

Moreover, a comparison of the four-qubit entanglement at $t\approx 2.85$ $\mu$s shown in Fig. \ref{f4}(b) with the tripartite entanglement at $t\approx 3.23$ $\mu$s in Fig. \ref{f2}(d) and the bipartite entanglement at $t= 5.325$ $\mu$s from Ref. \cite{ar1}, all for the same interaction strength $J$, demonstrates that four-qubit entanglement can be achieved more rapidly by slightly adjusting the resonant driving amplitude to $\Omega=1.598$ rad/$\mu s$. This fast entanglement generation can be attributed to higher-order EPs and rapid transitions of the qubits between different quantum states. These findings suggest that higher-order EPs in many-body quantum systems can be advantageous for generating multipartite entanglement. 

\subsection*{Hermitian Limit}

\begin{figure}
	\begin{center}
		\hspace{-2.75cm}(a)\hspace{4cm}{(b)}\\
		\includegraphics[width=0.5\textwidth]{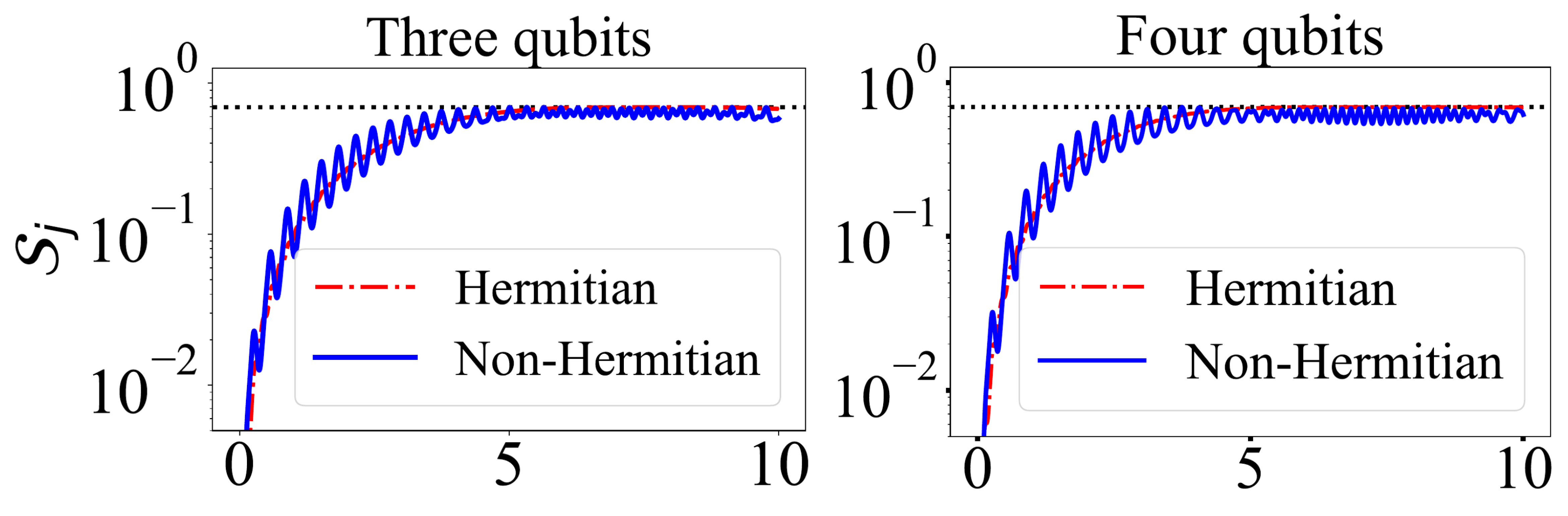}\\
		\hspace{-2.75cm}(c)\hspace{4cm}{(e)}\\
		\includegraphics[width=0.5\textwidth]{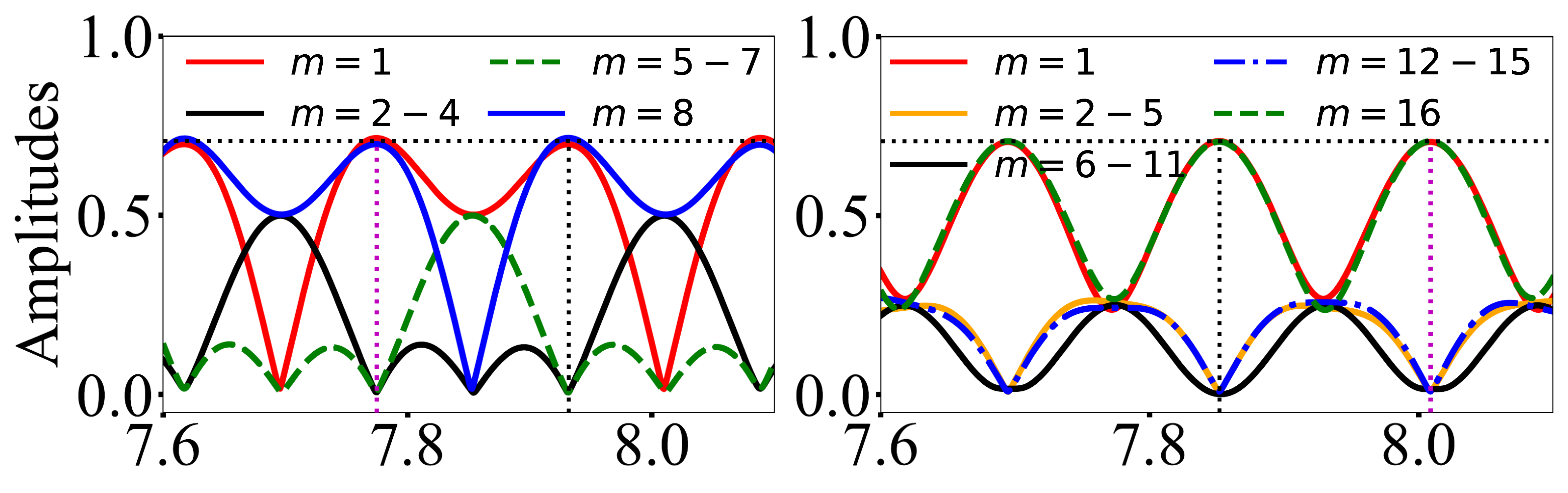}\\
		\hspace{-2.75cm}(d)\hspace{4cm}{(f)}\\
		\includegraphics[width=0.5\textwidth]{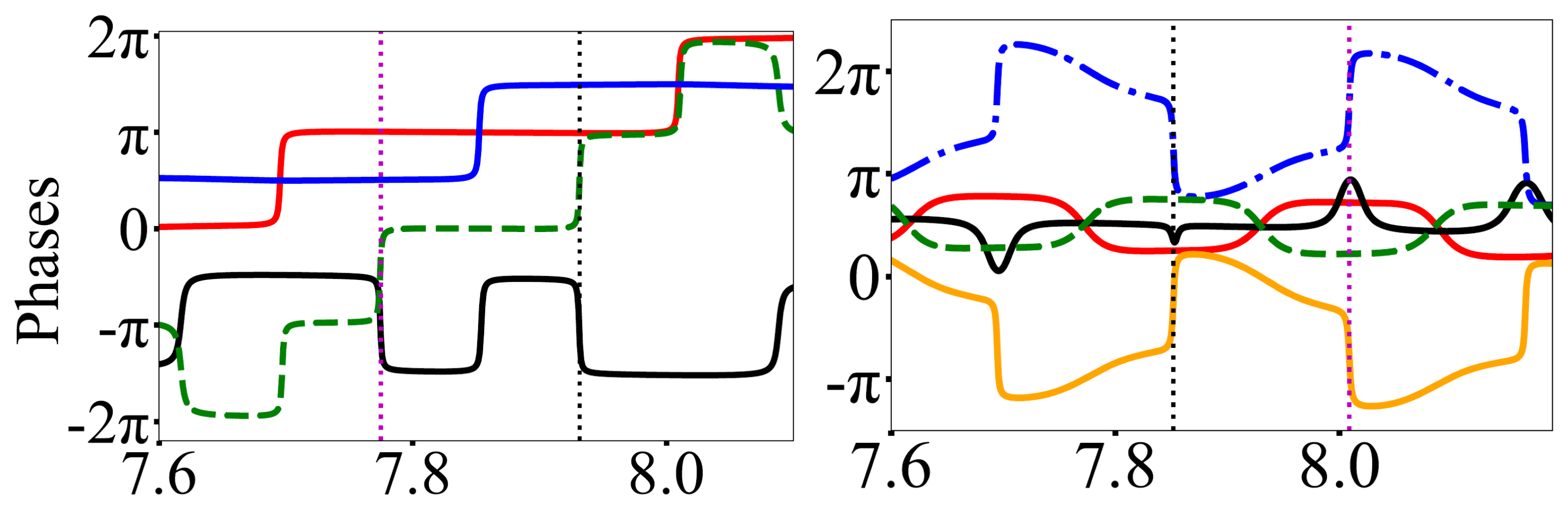}\\
		\hspace{-2.75cm}(g)\hspace{4cm}{(h)}\\
		\includegraphics[width=0.5\textwidth]{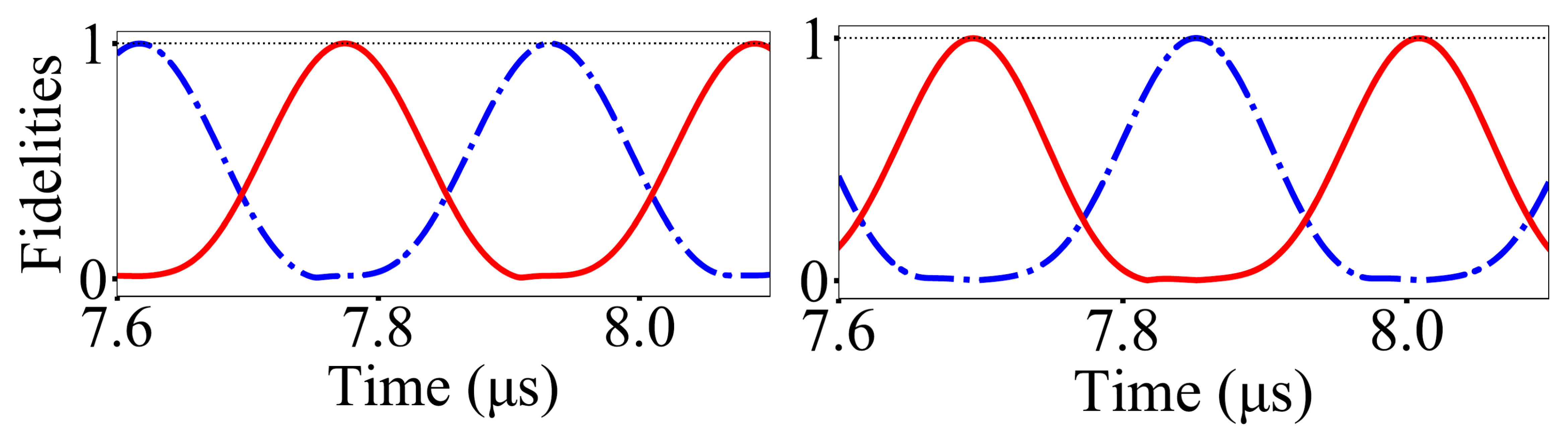}
		\caption{Dynamics of three and four qubits from the initial state $|\psi(0)\rangle=|f\rangle^{\otimes n}$. (a) and (b) Entanglement entropies for non-Hermitian qubits ($\gamma=6$ rad/$\mu s$, solid blue) and Hermitian qubits ($\gamma=0$, dash-dotted red). Panels (c) and (d) show dynamics of the amplitudes $|\alpha_m(t)|$ and phases ${\rm Arg}[\alpha_m(t)]$ of three Hermitian qubits of state $|\psi(t)\rangle=\sum_{m=1}^{8} \alpha_m(t) | \delta_m \rangle$, with $m=1$ (solid red), $m=2-4$ (solid black), $m=5-7$ (dashed green), and $m=8$ (solid blue). Panels (e) and (f) show the amplitudes $|\tilde{\alpha}_m(t)|$ and phases ${\rm Arg}[\tilde{\alpha}_m(t)]$ of four Hermitian qubits of state $| \tilde{\psi}(t) \rangle=\sum_{m=1}^{16} \tilde{\alpha}_m(t) | \tilde{\delta}_m \rangle$, with $m=1$ (solid red), $m=2-5$ (solid orange), $m=6-11$ (solid black), $m=12-15$ (dash-dotted blue), and $m=16$ (dashed green). Dotted black horizintal lines in (c) and (e) mark the amplitudes $(1/\sqrt{2})$ of the standard GHZ states, while dotted magenta and black vertical lines in (c)-(f) indicate the three- and four-qubit GHZ states $|\psi_{\pm}\rangle \approx e^{i\pi}(|fff\rangle + e^{\pm i\pi/2}|eee\rangle)/\sqrt{2}$ and $|\tilde{\psi}_{\mp}\rangle \approx e^{i\pi/4}(|ffff\rangle + e^{\mp i\pi/2}|eeee\rangle)/\sqrt{2}$. (g) Fidelities of the three-qubit GHZ states $|\psi_{+}\rangle$ (solid red) and $|\psi_{-}\rangle$ (dash-dotted blue). (h) Fidelities of the four-qubit GHZ states $|\tilde{\psi}_{-}\rangle$ (dash-dotted blue) and $|\tilde{\psi}_{+}\rangle$ (solid red). Other parameters are $\Delta=0$, $\Omega=10$ rad/$\mu s$, $J=0.4$ rad/$\mu s$.}\label{f5}
	\end{center} 
\end{figure} 

Up to now, we have concentrated on the dynamics of multipartite GHZ-class states near higher-order EPs, highlighting that non-Hermitian qubits can save thousands of microseconds compared to Hermitian qubits within the same parameter regime. The long time dynamics required for Hermitian qubits expose them to environment-induced decoherence effect, which is detrimental to entanglement generation. 

Alternatively, Hermitian qubits can generate three- and four-qubit entangled states on a timescale comparable to non-Hermitian qubits as demonstrated in Figs. \ref{f5}(a) and \ref{f5}(b). This is achieved by driving the qubits with a strong Rabi frequency $\Omega=10$ rad/$\mu$s and adjusting the coupling strength to $J=0.4$ rad/$\mu$s, while remaining in the weak coupling regime. In this parameter range, Hermitian effects are dominant, therefore the entanglement induced by non-Hermitian qubits showcases an oscillatory behavior on top of the Hermitian results. Additionally, this parameter regime not only speeds up entanglement generation in Hermitian qubits but also stabilizes the maximal entanglement for both initial states $|\psi(0)\rangle=2^{-n/2}(|f\rangle-i|e\rangle)^{\otimes n}$ and $|\psi(0)\rangle=|f\rangle^{\otimes n}$ after a few microseconds (supplementary note D for the details).

To explicitly identify the entangled states generated in stable regimes illustrated in Figs. \ref{f5}(a) and \ref{f5}(b), we examine the dynamics of qubit probability amplitudes and phases within the time interval $t \in [7.6$ $\mu$s, $8.1$ $\mu$s], as illustrated in Figs. \ref{f5}(c)-\ref{f5}(f). In both the three-qubit and four-qubit systems, the probability amplitudes oscillate over time, and the phases adjust accordingly to ensure the generation of a specific multipartite entangled state at a certain time. For example, when the probability amplitudes of the bases $|f\rangle^{\otimes n}$ and $|e\rangle^{\otimes n}$ are close to $1/\sqrt{2}$, and the amplitudes of the other degenerate bases become vanishing, this suggests the generation of GHZ states of the form $e^{i\theta_g} \left( |f\rangle^{\otimes n} + e^{i\theta_r} |e\rangle^{\otimes n} \right)/\sqrt{2}$ as illustrated by the vertical lines in Figs. \ref{f5}(c) and \ref{f5}(e). The global phase $\theta_g$ and the relative phase $\theta_r$ can be obtained from the phase dynamics shown in Figs. \ref{f5}(d) and \ref{f5}(f).

In particular, three-qubit GHZ states $|\psi_{\mp}\rangle\approx e^{i\pi}(|fff\rangle + e^{\mp i\pi/2}|eee\rangle)/\sqrt{2}$ can be generated at $t \approx 7.775$ $\mu\text{s}$ and $t \approx 7.932$ $\mu\text{s}$, respectively, as indicated in Figs. \ref{f5}(c) and \ref{f5}(d). In Fig. \ref{f5}(g), we depict their fidelities, defined as $|\langle{\rm GHZ}|\psi_{\pm}\rangle|$ with $|{\rm GHZ}\rangle=(|fff\rangle -i|eee\rangle)/\sqrt{2}$, and they are approximately $0.9997$ and $0.9998$ at the respective times. Similarly, four-qubit GHZ states $|\tilde{\psi}_{\pm}\rangle\approx e^{i\pi/4}(|ffff\rangle + e^{\pm i\pi/2}|eeee\rangle)/\sqrt{2}$ can be produced at $t \approx 7.852$ $\mu\text{s}$ and $t \approx 8.009$ $\mu\text{s}$, respectively, with the corresponding fidelities $|\langle\tilde{\rm GHZ}|\tilde{\psi}_{\pm}\rangle|\approx 0.9997$ demonstrated in Fig. \ref{f5}(h). 

Our system also hosts other maximally entangled GHZ classes $|\psi\rangle$ $\approx$ $0.5e^{-i\pi/2}(|ffe\rangle + |fef\rangle + |eff\rangle + e^{i\pi}|eee\rangle)$ and $|\psi\rangle$  $\approx$ $0.5(|fee\rangle + |efe\rangle + |eef\rangle + e^{i\pi}|fff\rangle)$ at $t \approx 7.715$ $\mu\text{s}$ and $t \approx 7.856$ $\mu\text{s}$, respectively. The former state is obtained by swapping $|fff\rangle$ in the standard GHZ state with the degenerate bases $\{|ffe\rangle$, $|fef\rangle$, $|eff\rangle\}$, while the latter state replaces $|eee\rangle$ with $\{|fee\rangle$, $|efe\rangle$, $|eef\rangle\}$. This observation indicates that tripartite entanglement remains unaffected by population exchange between $|fff\rangle$ and $\{|fee\rangle$, $|efe\rangle$, $|eef\rangle\}$, and between $|eee\rangle$ and $\{|ffe\rangle$, $|fef\rangle$, $|eff\rangle\}$, provided that an appropriate relative phase should be maintained to keep maximal entanglement. This property may be specific to odd numbers of qubits because, in the case of four-qubit entanglement, all probability amplitudes are non-vanishing except for the standard GHZ states. For example, four-qubit GHZ classes with nearly equal probability amplitudes are evident from Figs. \ref{f5}(e) and \ref{f5}(f) at $t \approx 7.782$ $\mu\text{s}$ and $t \approx 7.945$ $\mu\text{s}$. Here, entanglement primarily arises from phase buildup, similar to the two-qubit scenario \cite{ar1}.

\subsection*{Discussions}
In conclusion, we generate three- and four-qubit entanglement near higher-order EPs that correspond to the dimension of their Hilbert spaces. We also present entangled states generation in the strongly-driven Hermitian qubits with high fidelity to the GHZ states as a comparison to the non-Hermitian cases. Our results show that non-Hermitian qubits can simultaneously save significant driving energy and accelerate multipartite entanglement generation. In particular, they can reduce the required multiparty entanglement generating time by thousands of microseconds compared to Hermitian qubits operating in the same parameter regimes. This advantage is attributed to the ultra-high sensitivity of non-Hermitian qubits to parameter changes and initial conditions near EPs. Notably, our result points out that higher-order EPs in many-body quantum systems can be beneficial for multipartite entanglement generation. Furthermore, these EPs allow a design of specific driving protocols and inter-qubit couplings to reach desired multipartite entangled states at precise times, which is crucial for practical quantum technologies and fundamental studies. Non-Hermitian qubits thus exhibit unique multipartite entanglement dynamics not observed in Hermitian systems. 

Furthermore, we show that achieving entanglement with Hermitian qubits on the same timescale as non-Hermitian qubits requires strong resonant driving, which also stabilizes maximal entanglement for several time periods. Observations of the dynamics of probability amplitudes and phases in stable entangled regimes reveals fundamental mechanisms that lead to GHZ states in both odd and even numbers of qubits.

Although our work concentrates on GHZ classes, it is interesting to see whether non-Hermitian system can generate W states and other distinct classes of multiply entangled states e.g., graph states \cite{ar19, ar31}. Our result can naturally be extended to scalable GHZ states and realized through parametrically driven and Josephson-based quantum state routers \cite{ar27} and superconducting processors with all-to-all coupling \cite{ar28}.
\section*{Methods}
The joint evolution of the qubits from an arbitrary initial state $|\psi(0)\rangle$ is given by ($\hbar=1$)   
\begin{eqnarray}
|\psi(t)\rangle&=&\frac{e^{-i\hat{H}t}|\psi(0)\rangle}{\sqrt{\langle\psi(0)|e^{i\hat{H^\dag}t}e^{-i\hat{H}t}|\psi(0)\rangle}}, \label{meth}
\end{eqnarray}
where $\hat{H}$ is the Hamiltonian defined in Eq. (1).

We can also rewrite the state in Eq. (\ref{meth}) as 
\begin{eqnarray}
|\psi(t)\rangle=\sum^{2^n}_{m=1}\langle\tilde{\phi}_m|\psi(0)\rangle e^{-iE_mt}|\phi_m\rangle,
\end{eqnarray}
where the summation index $2^n$ denotes the dimension of the Hilbert space spanned by $n$ qubits, $E_m$ represents the complex eigenvalues, while $|\phi_m\rangle$ and $\langle\tilde{\phi}_m|$ are the normalized right and left biorthogonal eigenvectors. The state can also be expanded in terms of the computational bases. For example, the three-qubit state is given by
\begin{eqnarray}
|\psi(t)\rangle=\sum^{8}_{m=1}\alpha_m(t)|\delta_m\rangle,
\end{eqnarray}
where $\alpha_m(t)$ indicates the complex probability amplitudes, and $|\delta_m\rangle$ is the basis states. These bases include $|fff\rangle$ for $m=1$, a set of degenerate bases $\{|ffe\rangle, |fef\rangle$, $|eff\rangle\}$ for $m=2-4$, another set of degenerate bases $\{|fee\rangle, |efe\rangle$, $|eef\rangle\}$ for $m=5-7$, and $|eee\rangle$ for $m=8$. 

Similarly, the four-qubit state can be expressed as
\begin{eqnarray}
|{\tilde\psi(t)}\rangle=\sum^{16}_{m=1}{\tilde\alpha_m(t)}|{\tilde\delta_m}\rangle,
\end{eqnarray}
with the new probability amplitudes ${\tilde\alpha(t)}$ and bases $|\tilde\delta_m\rangle$. Four qubits have sixteen computational bases, which can be grouped into five as $|ffff\rangle$ for $m=1$, the first set of degenerate basis $\{|fffe\rangle, ..., |efff\rangle\}$ for $m=2-5$, the second set of degenerate basis $\{|ffee\rangle,..., |eeff\rangle\}$ for $m=6-11$, the third set of degenerate basis $\{|feee\rangle,..., |eeef\rangle\}$ for $m=12-15$, and lastly $|eeee\rangle$ for $m=16$ (see Figs. \ref{f5}(e) and \ref{f5}(f)).

The probability amplitudes and phases discussed in the main text are computed using the QuTiP software packages \cite{ar15}. This approach enables us to determine the states of non-Hermitian qubits at any given time by specifying their probability amplitudes and phase factors. For instance, standard three-qubit and four-qubit GHZ states are obtained when the amplitudes of the basis states $|f\rangle^{\otimes n}$ and $|e\rangle^{\otimes n}$ are exactly $1/\sqrt{2}$, while all other probability amplitudes are vanishing, as illustrated in Figs. \ref{f5}(c)-\ref{f5}(f) (see also Figs. \ref{f2}(a)-\ref{f2}(c)).

Entanglement entropy quantifies the amount of information needed to fully describe the state of one part of a quantum system when the other parts are inaccessible \cite{ar7}. For example, we can divide the three-qubit system into subsystem A consisting reduced qubit $j$ and subsystem B consisting the remaining two qubits. Quantum correlation between the subsystems can be evaluated by the entanglement entropy (equal for both subsystems) of the reduced qubit as ${\cal S}_{j}=-{\rm Tr}[\hat{\rho}_{j}(t) \log \hat{\rho}_{j}(t)],$ where the reduced density matrices $\hat{\rho}_{j}(t)$ of each qubit are given by
\begin{widetext}
\begin{eqnarray}
\hat{\rho}_1(t)&=& 
\begin{pmatrix}
|\alpha_1(t)|^2 + |\alpha_2(t)|^2 + |\alpha_3(t)|^2 + |\alpha_5(t)|^2 & \alpha_1(t)\alpha_4^*(t) + \alpha_2(t)\alpha_6^*(t) + \alpha_3(t)\alpha_7^*(t) + \alpha_5(t)\alpha_8^*(t)\\ 
\alpha_4(t)\alpha_1^*(t) + \alpha_6(t)\alpha_2^*(t) + \alpha_7(t)\alpha_3^*(t) + \alpha_8(t)\alpha_5^*(t) & |\alpha_4(t)|^2 + |\alpha_6(t)|^2 + |\alpha_7(t)|^2 + |\alpha_8(t)|^2
\end{pmatrix},\nonumber\\
\hat{\rho}_2(t)&=& 
\begin{pmatrix}
|\alpha_1(t)|^2 + |\alpha_2(t)|^2 + |\alpha_4(t)|^2 + |\alpha_6(t)|^2 & \alpha_1(t)\alpha_3^*(t) + \alpha_2(t)\alpha_5^*(t) + \alpha_4(t)\alpha_7^*(t) + \alpha_6(t)\alpha_8^*(t)\\ 
\alpha_3(t)\alpha_1^*(t) + \alpha_5(t)\alpha_2^*(t) + \alpha_7(t)\alpha_4^*(t) + \alpha_8(t)\alpha_6^*(t) & |\alpha_3(t)|^2 + |\alpha_5(t)|^2 + |\alpha_7(t)|^2 + |\alpha_8(t)|^2
\end{pmatrix},\nonumber\\
\hat{\rho}_3(t)&=&
\begin{pmatrix}
|\alpha_1(t)|^2 + |\alpha_3(t)|^2 + |\alpha_4(t)|^2 + |\alpha_7(t)|^2 & \alpha_1(t)\alpha_2^*(t) + \alpha_3(t)\alpha_5^*(t) + \alpha_4(t)\alpha_6^*(t) + \alpha_7(t)\alpha_8^*(t)\\ 
\alpha_2(t)\alpha_1^*(t) + \alpha_5(t)\alpha_3^*(t) + \alpha_6(t)\alpha_4^*(t) + \alpha_8(t)\alpha_7^*(t) & |\alpha_2(t)|^2 + |\alpha_5(t)|^2 + |\alpha_6(t)|^2 + |\alpha_8(t)|^2
\end{pmatrix}. \nonumber\\\label{m1}
\end{eqnarray}
\end{widetext}

Donating the reduced density matrices of Eq. (\ref{m1}) by 
\begin{eqnarray}
\hat{\rho}_{j}(t)=
\begin{pmatrix}
\rho^{j}_{xx} & \rho^{j}_{xy}\\ 
\rho^{j}_{yx} & \rho^{j}_{yy}
\end{pmatrix}, \label{mm0}
\end{eqnarray}
and noting $\rho^{j}_{xx}+\rho^{j}_{yy}=1$,
the entanglement entropy of qubit $j$ can further reduce to 
\begin{eqnarray}
{\cal S}_{j}&=&-\lambda^{j}_- \log \lambda^{j}_- - \lambda^{j}_+ \log \lambda^{j}_+,
\end{eqnarray}
where $\lambda^{j}_{\pm}=\frac{1}{2}\pm\frac{1}{2}\sqrt{(\rho^{j}_{xx}-\rho^{j}_{yy})^2+4\rho^{j}_{xy}\rho^{j}_{yx}}$ are the eigenvalues of Eq. \ref{mm0}.

The entanglement entropy of the reduced qubits takes on distinct values for product states, GHZ states, and W states. This distinction not only allows us to differentiate between product and entangled states but also indicates that GHZ states capture greater quantum correlations compared to W states. For GHZ states, the eigenvalues $\lambda^{j}_{+}$ and $\lambda^{j}_{-}$ are both $1/2$. In contrast, for product states, either $\lambda^{j}_{+}$ or $\lambda^{j}_{-}$ is zero, while the remaining eigenvalue is unity. For instance, in the case of the initial coherent superposition considered in the results section, we have $\lambda^{j}_{+}=0$ and $\lambda^{j}_{-}=1$, leading to ${\cal S}_{j}=0$.

Consequently, the entanglement entropy of the reduced qubits ranges from ${\cal S}_{1}={\cal S}_{2}={\cal S}_{3}=0$ for product states to ${\cal S}_{1}={\cal S}_{2}={\cal S}_{3}=\log{2}$ for GHZ states. For W states, it takes a specific value ${\cal S}_{1}={\cal S}_{2}={\cal S}_{3}=\log{3}-(2/3)\log{2}$, which falls within the interval $0\leq{\cal S}_{j}\leq\log{2}$. The same idea can apply to large number qubits. For instance, entanglement entropy has successfully identified the four-qubit GHZ state with a fidelity of approximately $0.999$ in the Hermitian limit (see Fig. \ref{f5}(b) and Fig. \ref{f5}(h)). 

Furthermore, we use three tangle, often known as the residual three-qubit entanglement measure specifically designed to quantify the amount of genuine entanglement shared among all three qubits, excluding any bipartite entanglement contributions. It thus captures the essence of entanglement monogamy by showing that the entanglement is distributed among the three qubits in a constrained way as \cite{ar10} 
\begin{eqnarray}
\tau_{123}={\cal C}^2_{1(23)}-{\cal C}^2_{12}-{\cal C}^2_{13}, \label{m10}                                                                                                                                                                    \end{eqnarray}
where ${\cal C}_{12}$ and ${\cal C}_{13}$ represent pairwise concurrences between two qubits \cite{ar20}. For instance, the concurrence ${\cal C}_{12}$ is defined as \cite{ar20} ${\cal C}_{12}=\text{max}(0, \lambda_{1}-\lambda_{2}-\lambda_{3}-\lambda_{4}),$ where $\lambda_{1},$ $\lambda_{2},$ $\lambda_{3}$ and $\lambda_{4}$ are the eigenvalues of the Hermitian matrix $\hat R_{12}=\sqrt{\sqrt{\hat\rho_{12}(t)}\tilde{\hat\rho}_{12}(t)\sqrt{\hat\rho_{12}(t)}}$, with $\tilde{\hat\rho}_{12}(t)=(\sigma_y\otimes\sigma_y)\hat\rho^{\ast}_{12}(t)(\sigma_y\otimes\sigma_y)$, and $\hat\rho_{12}(t)=\text{Tr}_{3}[|\psi(t)\rangle\langle\psi(t)|].$ The eigenvalues should be ordered as $\lambda_{1}\geq\lambda_{2}\geq\lambda_{3}\geq\lambda_{4}$. 

Moreover, ${\cal C}_{1(23)}$ in Eq. \ref{m10} represents a bi-partition concurrence calculated from the purity ${P}_{1}={\rm Tr}[\hat\rho^2_1(t)]$ of the first qubit as \cite{ar9} ${\cal C}_{1(23)}=\sqrt{2-2{P}_{1}}$. The three-tangle is invariant under permutations of the qubits \cite{ar10}; thus, rearrangement of the qubits in any order does not change the value of the three-tangle. 

\nocite{*}
\bibliography{references} 

\providecommand{\noopsort}[1]{}\providecommand{\singleletter}[1]{#1}%
\begin{thebibliography}{72}%
\makeatletter
\providecommand \@ifxundefined [1]{%
 \@ifx{#1\undefined}
}%
\providecommand \@ifnum [1]{%
 \ifnum #1\expandafter \@firstoftwo
 \else \expandafter \@secondoftwo
 \fi
}%
\providecommand \@ifx [1]{%
 \ifx #1\expandafter \@firstoftwo
 \else \expandafter \@secondoftwo
 \fi
}%
\providecommand \natexlab [1]{#1}%
\providecommand \enquote  [1]{``#1''}%
\providecommand \bibnamefont  [1]{#1}%
\providecommand \bibfnamefont [1]{#1}%
\providecommand \citenamefont [1]{#1}%
\providecommand \href@noop [0]{\@secondoftwo}%
\providecommand \href [0]{\begingroup \@sanitize@url \@href}%
\providecommand \@href[1]{\@@startlink{#1}\@@href}%
\providecommand \@@href[1]{\endgroup#1\@@endlink}%
\providecommand \@sanitize@url [0]{\catcode `\\12\catcode `\$12\catcode
  `\&12\catcode `\#12\catcode `\^12\catcode `\_12\catcode `\%12\relax}%
\providecommand \@@startlink[1]{}%
\providecommand \@@endlink[0]{}%
\providecommand \url  [0]{\begingroup\@sanitize@url \@url }%
\providecommand \@url [1]{\endgroup\@href {#1}{\urlprefix }}%
\providecommand \urlprefix  [0]{URL }%
\providecommand \Eprint [0]{\href }%
\providecommand \doibase [0]{https://doi.org/}%
\providecommand \selectlanguage [0]{\@gobble}%
\providecommand \bibinfo  [0]{\@secondoftwo}%
\providecommand \bibfield  [0]{\@secondoftwo}%
\providecommand \translation [1]{[#1]}%
\providecommand \BibitemOpen [0]{}%
\providecommand \bibitemStop [0]{}%
\providecommand \bibitemNoStop [0]{.\EOS\space}%
\providecommand \EOS [0]{\spacefactor3000\relax}%
\providecommand \BibitemShut  [1]{\csname bibitem#1\endcsname}%
\let\auto@bib@innerbib\@empty
\bibitem [{\citenamefont {Sakurai}\ and\ \citenamefont
  {Napolitano}(2020)}]{ar00}%
  \BibitemOpen
  \bibfield  {author} {\bibinfo {author} {\bibfnamefont {J.~J.}\ \bibnamefont
  {Sakurai}}\ and\ \bibinfo {author} {\bibfnamefont {J.}~\bibnamefont
  {Napolitano}},\ }\href@noop {} {\emph {\bibinfo {title} {Modern quantum
  mechanics}}}\ (\bibinfo  {publisher} {Cambridge University Press},\ \bibinfo
  {year} {2020})\BibitemShut {NoStop}%
\bibitem [{\citenamefont {Breuer}\ and\ \citenamefont
  {Petruccione}(2002)}]{ar01}%
  \BibitemOpen
  \bibfield  {author} {\bibinfo {author} {\bibfnamefont {H.-P.}\ \bibnamefont
  {Breuer}}\ and\ \bibinfo {author} {\bibfnamefont {F.}~\bibnamefont
  {Petruccione}},\ }\href@noop {} {\emph {\bibinfo {title} {The theory of open
  quantum systems}}}\ (\bibinfo  {publisher} {Oxford University Press, USA},\
  \bibinfo {year} {2002})\BibitemShut {NoStop}%
\bibitem [{\citenamefont {Minganti}\ \emph {et~al.}(2019)\citenamefont
  {Minganti}, \citenamefont {Miranowicz}, \citenamefont {Chhajlany},\ and\
  \citenamefont {Nori}}]{ar02}%
  \BibitemOpen
  \bibfield  {author} {\bibinfo {author} {\bibfnamefont {F.}~\bibnamefont
  {Minganti}}, \bibinfo {author} {\bibfnamefont {A.}~\bibnamefont
  {Miranowicz}}, \bibinfo {author} {\bibfnamefont {R.~W.}\ \bibnamefont
  {Chhajlany}},\ and\ \bibinfo {author} {\bibfnamefont {F.}~\bibnamefont
  {Nori}},\ }\bibfield  {title} {\bibinfo {title} {Quantum exceptional points
  of non-hermitian hamiltonians and liouvillians: The effects of quantum
  jumps},\ }\href@noop {} {\bibfield  {journal} {\bibinfo  {journal} {Physical
  Review A}\ }\textbf {\bibinfo {volume} {100}},\ \bibinfo {pages} {062131}
  (\bibinfo {year} {2019})}\BibitemShut {NoStop}%
\bibitem [{\citenamefont {Bender}\ and\ \citenamefont
  {Boettcher}(1998)}]{ar24}%
  \BibitemOpen
  \bibfield  {author} {\bibinfo {author} {\bibfnamefont {C.~M.}\ \bibnamefont
  {Bender}}\ and\ \bibinfo {author} {\bibfnamefont {S.}~\bibnamefont
  {Boettcher}},\ }\bibfield  {title} {\bibinfo {title} {Real spectra in
  non-hermitian hamiltonians having p t symmetry},\ }\href@noop {} {\bibfield
  {journal} {\bibinfo  {journal} {Physical Review Letters}\ }\textbf {\bibinfo
  {volume} {80}},\ \bibinfo {pages} {5243} (\bibinfo {year}
  {1998})}\BibitemShut {NoStop}%
\bibitem [{\citenamefont {Okuma}\ and\ \citenamefont {Sato}(2023)}]{ar23}%
  \BibitemOpen
  \bibfield  {author} {\bibinfo {author} {\bibfnamefont {N.}~\bibnamefont
  {Okuma}}\ and\ \bibinfo {author} {\bibfnamefont {M.}~\bibnamefont {Sato}},\
  }\bibfield  {title} {\bibinfo {title} {Non-hermitian topological phenomena: A
  review},\ }\href@noop {} {\bibfield  {journal} {\bibinfo  {journal} {Annual
  Review of Condensed Matter Physics}\ }\textbf {\bibinfo {volume} {14}},\
  \bibinfo {pages} {83} (\bibinfo {year} {2023})}\BibitemShut {NoStop}%
\bibitem [{\citenamefont {Li}\ \emph {et~al.}(2023)\citenamefont {Li},
  \citenamefont {Chen}, \citenamefont {Abbasi}, \citenamefont {Murch},\ and\
  \citenamefont {Whaley}}]{ar1}%
  \BibitemOpen
  \bibfield  {author} {\bibinfo {author} {\bibfnamefont {Z.-Z.}\ \bibnamefont
  {Li}}, \bibinfo {author} {\bibfnamefont {W.}~\bibnamefont {Chen}}, \bibinfo
  {author} {\bibfnamefont {M.}~\bibnamefont {Abbasi}}, \bibinfo {author}
  {\bibfnamefont {K.~W.}\ \bibnamefont {Murch}},\ and\ \bibinfo {author}
  {\bibfnamefont {K.~B.}\ \bibnamefont {Whaley}},\ }\bibfield  {title}
  {\bibinfo {title} {Speeding up entanglement generation by proximity to
  higher-order exceptional points},\ }\href@noop {} {\bibfield  {journal}
  {\bibinfo  {journal} {Physical Review Letters}\ }\textbf {\bibinfo {volume}
  {131}},\ \bibinfo {pages} {100202} (\bibinfo {year} {2023})}\BibitemShut
  {NoStop}%
\bibitem [{\citenamefont {Naghiloo}\ \emph {et~al.}(2019)\citenamefont
  {Naghiloo}, \citenamefont {Abbasi}, \citenamefont {Joglekar},\ and\
  \citenamefont {Murch}}]{ar2}%
  \BibitemOpen
  \bibfield  {author} {\bibinfo {author} {\bibfnamefont {M.}~\bibnamefont
  {Naghiloo}}, \bibinfo {author} {\bibfnamefont {M.}~\bibnamefont {Abbasi}},
  \bibinfo {author} {\bibfnamefont {Y.~N.}\ \bibnamefont {Joglekar}},\ and\
  \bibinfo {author} {\bibfnamefont {K.}~\bibnamefont {Murch}},\ }\bibfield
  {title} {\bibinfo {title} {Quantum state tomography across the exceptional
  point in a single dissipative qubit},\ }\href@noop {} {\bibfield  {journal}
  {\bibinfo  {journal} {Nature Physics}\ }\textbf {\bibinfo {volume} {15}},\
  \bibinfo {pages} {1232} (\bibinfo {year} {2019})}\BibitemShut {NoStop}%
\bibitem [{\citenamefont {Chen}\ \emph {et~al.}(2021)\citenamefont {Chen},
  \citenamefont {Abbasi}, \citenamefont {Joglekar},\ and\ \citenamefont
  {Murch}}]{ar3}%
  \BibitemOpen
  \bibfield  {author} {\bibinfo {author} {\bibfnamefont {W.}~\bibnamefont
  {Chen}}, \bibinfo {author} {\bibfnamefont {M.}~\bibnamefont {Abbasi}},
  \bibinfo {author} {\bibfnamefont {Y.~N.}\ \bibnamefont {Joglekar}},\ and\
  \bibinfo {author} {\bibfnamefont {K.~W.}\ \bibnamefont {Murch}},\ }\bibfield
  {title} {\bibinfo {title} {Quantum jumps in the non-hermitian dynamics of a
  superconducting qubit},\ }\href@noop {} {\bibfield  {journal} {\bibinfo
  {journal} {Physical Review Letters}\ }\textbf {\bibinfo {volume} {127}},\
  \bibinfo {pages} {140504} (\bibinfo {year} {2021})}\BibitemShut {NoStop}%
\bibitem [{\citenamefont {Ashida}\ \emph {et~al.}(2020)\citenamefont {Ashida},
  \citenamefont {Gong},\ and\ \citenamefont {Ueda}}]{ar21}%
  \BibitemOpen
  \bibfield  {author} {\bibinfo {author} {\bibfnamefont {Y.}~\bibnamefont
  {Ashida}}, \bibinfo {author} {\bibfnamefont {Z.}~\bibnamefont {Gong}},\ and\
  \bibinfo {author} {\bibfnamefont {M.}~\bibnamefont {Ueda}},\ }\bibfield
  {title} {\bibinfo {title} {Non-hermitian physics},\ }\href@noop {} {\bibfield
   {journal} {\bibinfo  {journal} {Advances in Physics}\ }\textbf {\bibinfo
  {volume} {69}},\ \bibinfo {pages} {249} (\bibinfo {year} {2020})}\BibitemShut
  {NoStop}%
\bibitem [{\citenamefont {Gong}\ \emph {et~al.}(2018)\citenamefont {Gong},
  \citenamefont {Ashida}, \citenamefont {Kawabata}, \citenamefont {Takasan},
  \citenamefont {Higashikawa},\ and\ \citenamefont {Ueda}}]{ar515}%
  \BibitemOpen
  \bibfield  {author} {\bibinfo {author} {\bibfnamefont {Z.}~\bibnamefont
  {Gong}}, \bibinfo {author} {\bibfnamefont {Y.}~\bibnamefont {Ashida}},
  \bibinfo {author} {\bibfnamefont {K.}~\bibnamefont {Kawabata}}, \bibinfo
  {author} {\bibfnamefont {K.}~\bibnamefont {Takasan}}, \bibinfo {author}
  {\bibfnamefont {S.}~\bibnamefont {Higashikawa}},\ and\ \bibinfo {author}
  {\bibfnamefont {M.}~\bibnamefont {Ueda}},\ }\bibfield  {title} {\bibinfo
  {title} {Topological phases of non-hermitian systems},\ }\href@noop {}
  {\bibfield  {journal} {\bibinfo  {journal} {Physical Review X}\ }\textbf
  {\bibinfo {volume} {8}},\ \bibinfo {pages} {031079} (\bibinfo {year}
  {2018})}\BibitemShut {NoStop}%
\bibitem [{\citenamefont {Kawabata}\ \emph {et~al.}(2019)\citenamefont
  {Kawabata}, \citenamefont {Shiozaki}, \citenamefont {Ueda},\ and\
  \citenamefont {Sato}}]{ar25}%
  \BibitemOpen
  \bibfield  {author} {\bibinfo {author} {\bibfnamefont {K.}~\bibnamefont
  {Kawabata}}, \bibinfo {author} {\bibfnamefont {K.}~\bibnamefont {Shiozaki}},
  \bibinfo {author} {\bibfnamefont {M.}~\bibnamefont {Ueda}},\ and\ \bibinfo
  {author} {\bibfnamefont {M.}~\bibnamefont {Sato}},\ }\bibfield  {title}
  {\bibinfo {title} {Symmetry and topology in non-hermitian physics},\
  }\href@noop {} {\bibfield  {journal} {\bibinfo  {journal} {Physical Review
  X}\ }\textbf {\bibinfo {volume} {9}},\ \bibinfo {pages} {041015} (\bibinfo
  {year} {2019})}\BibitemShut {NoStop}%
\bibitem [{\citenamefont {{Lin}}\ \emph {et~al.}(2024)\citenamefont {{Lin}},
  \citenamefont {{Kuo}}, \citenamefont {{Lambert}}, \citenamefont
  {{Miranowicz}}, \citenamefont {{Nori}},\ and\ \citenamefont
  {{Chen}}}]{Lin2024arxiv}%
  \BibitemOpen
  \bibfield  {author} {\bibinfo {author} {\bibfnamefont {J.-D.}\ \bibnamefont
  {{Lin}}}, \bibinfo {author} {\bibfnamefont {P.-C.}\ \bibnamefont {{Kuo}}},
  \bibinfo {author} {\bibfnamefont {N.}~\bibnamefont {{Lambert}}}, \bibinfo
  {author} {\bibfnamefont {A.}~\bibnamefont {{Miranowicz}}}, \bibinfo {author}
  {\bibfnamefont {F.}~\bibnamefont {{Nori}}},\ and\ \bibinfo {author}
  {\bibfnamefont {Y.-N.}\ \bibnamefont {{Chen}}},\ }\bibfield  {title}
  {\bibinfo {title} {{Non-Markovian Quantum Exceptional Points}},\ }\href
  {https://doi.org/10.48550/arXiv.2406.18362} {\bibfield  {journal} {\bibinfo
  {journal} {arXiv e-prints}\ ,\ \bibinfo {eid} {arXiv:2406.18362}} (\bibinfo
  {year} {2024})},\ \Eprint {https://arxiv.org/abs/2406.18362}
  {arXiv:2406.18362 [quant-ph]} \BibitemShut {NoStop}%
\bibitem [{\citenamefont {Bender}(2007)}]{ar5}%
  \BibitemOpen
  \bibfield  {author} {\bibinfo {author} {\bibfnamefont {C.~M.}\ \bibnamefont
  {Bender}},\ }\bibfield  {title} {\bibinfo {title} {Making sense of
  non-hermitian hamiltonians},\ }\href@noop {} {\bibfield  {journal} {\bibinfo
  {journal} {Reports on Progress in Physics}\ }\textbf {\bibinfo {volume}
  {70}},\ \bibinfo {pages} {947} (\bibinfo {year} {2007})}\BibitemShut
  {NoStop}%
\bibitem [{\citenamefont {Mandal}\ and\ \citenamefont
  {Bergholtz}(2021)}]{ar03}%
  \BibitemOpen
  \bibfield  {author} {\bibinfo {author} {\bibfnamefont {I.}~\bibnamefont
  {Mandal}}\ and\ \bibinfo {author} {\bibfnamefont {E.~J.}\ \bibnamefont
  {Bergholtz}},\ }\bibfield  {title} {\bibinfo {title} {Symmetry and
  higher-order exceptional points},\ }\href@noop {} {\bibfield  {journal}
  {\bibinfo  {journal} {Physical Review Letters}\ }\textbf {\bibinfo {volume}
  {127}},\ \bibinfo {pages} {186601} (\bibinfo {year} {2021})}\BibitemShut
  {NoStop}%
\bibitem [{\citenamefont {Wiersig}(2014)}]{ar501}%
  \BibitemOpen
  \bibfield  {author} {\bibinfo {author} {\bibfnamefont {J.}~\bibnamefont
  {Wiersig}},\ }\bibfield  {title} {\bibinfo {title} {Enhancing the sensitivity
  of frequency and energy splitting detection by using exceptional points:
  application to microcavity sensors for single-particle detection},\
  }\href@noop {} {\bibfield  {journal} {\bibinfo  {journal} {Physical Review
  Letters}\ }\textbf {\bibinfo {volume} {112}},\ \bibinfo {pages} {203901}
  (\bibinfo {year} {2014})}\BibitemShut {NoStop}%
\bibitem [{\citenamefont {Chen}\ \emph {et~al.}(2017)\citenamefont {Chen},
  \citenamefont {Kaya~{\"O}zdemir}, \citenamefont {Zhao}, \citenamefont
  {Wiersig},\ and\ \citenamefont {Yang}}]{ar502}%
  \BibitemOpen
  \bibfield  {author} {\bibinfo {author} {\bibfnamefont {W.}~\bibnamefont
  {Chen}}, \bibinfo {author} {\bibfnamefont {{\c{S}}.}~\bibnamefont
  {Kaya~{\"O}zdemir}}, \bibinfo {author} {\bibfnamefont {G.}~\bibnamefont
  {Zhao}}, \bibinfo {author} {\bibfnamefont {J.}~\bibnamefont {Wiersig}},\ and\
  \bibinfo {author} {\bibfnamefont {L.}~\bibnamefont {Yang}},\ }\bibfield
  {title} {\bibinfo {title} {Exceptional points enhance sensing in an optical
  microcavity},\ }\href@noop {} {\bibfield  {journal} {\bibinfo  {journal}
  {Nature}\ }\textbf {\bibinfo {volume} {548}},\ \bibinfo {pages} {192}
  (\bibinfo {year} {2017})}\BibitemShut {NoStop}%
\bibitem [{\citenamefont {Miri}\ and\ \citenamefont {Alu}(2019)}]{ar503}%
  \BibitemOpen
  \bibfield  {author} {\bibinfo {author} {\bibfnamefont {M.-A.}\ \bibnamefont
  {Miri}}\ and\ \bibinfo {author} {\bibfnamefont {A.}~\bibnamefont {Alu}},\
  }\bibfield  {title} {\bibinfo {title} {Exceptional points in optics and
  photonics},\ }\href@noop {} {\bibfield  {journal} {\bibinfo  {journal}
  {Science}\ }\textbf {\bibinfo {volume} {363}},\ \bibinfo {pages} {eaar7709}
  (\bibinfo {year} {2019})}\BibitemShut {NoStop}%
\bibitem [{\citenamefont {R{\"u}ter}\ \emph {et~al.}(2010)\citenamefont
  {R{\"u}ter}, \citenamefont {Makris}, \citenamefont {El-Ganainy},
  \citenamefont {Christodoulides}, \citenamefont {Segev},\ and\ \citenamefont
  {Kip}}]{ar504}%
  \BibitemOpen
  \bibfield  {author} {\bibinfo {author} {\bibfnamefont {C.~E.}\ \bibnamefont
  {R{\"u}ter}}, \bibinfo {author} {\bibfnamefont {K.~G.}\ \bibnamefont
  {Makris}}, \bibinfo {author} {\bibfnamefont {R.}~\bibnamefont {El-Ganainy}},
  \bibinfo {author} {\bibfnamefont {D.~N.}\ \bibnamefont {Christodoulides}},
  \bibinfo {author} {\bibfnamefont {M.}~\bibnamefont {Segev}},\ and\ \bibinfo
  {author} {\bibfnamefont {D.}~\bibnamefont {Kip}},\ }\bibfield  {title}
  {\bibinfo {title} {Observation of parity--time symmetry in optics},\
  }\href@noop {} {\bibfield  {journal} {\bibinfo  {journal} {Nature physics}\
  }\textbf {\bibinfo {volume} {6}},\ \bibinfo {pages} {192} (\bibinfo {year}
  {2010})}\BibitemShut {NoStop}%
\bibitem [{\citenamefont {Choi}\ \emph {et~al.}(2017)\citenamefont {Choi},
  \citenamefont {Hahn}, \citenamefont {Yoon}, \citenamefont {Song},\ and\
  \citenamefont {Berini}}]{ar505}%
  \BibitemOpen
  \bibfield  {author} {\bibinfo {author} {\bibfnamefont {Y.}~\bibnamefont
  {Choi}}, \bibinfo {author} {\bibfnamefont {C.}~\bibnamefont {Hahn}}, \bibinfo
  {author} {\bibfnamefont {J.~W.}\ \bibnamefont {Yoon}}, \bibinfo {author}
  {\bibfnamefont {S.~H.}\ \bibnamefont {Song}},\ and\ \bibinfo {author}
  {\bibfnamefont {P.}~\bibnamefont {Berini}},\ }\bibfield  {title} {\bibinfo
  {title} {Extremely broadband, on-chip optical nonreciprocity enabled by
  mimicking nonlinear anti-adiabatic quantum jumps near exceptional points},\
  }\href@noop {} {\bibfield  {journal} {\bibinfo  {journal} {Nature
  Communications}\ }\textbf {\bibinfo {volume} {8}},\ \bibinfo {pages} {14154}
  (\bibinfo {year} {2017})}\BibitemShut {NoStop}%
\bibitem [{\citenamefont {Hodaei}\ \emph {et~al.}(2014)\citenamefont {Hodaei},
  \citenamefont {Miri}, \citenamefont {Heinrich}, \citenamefont
  {Christodoulides},\ and\ \citenamefont {Khajavikhan}}]{ar506}%
  \BibitemOpen
  \bibfield  {author} {\bibinfo {author} {\bibfnamefont {H.}~\bibnamefont
  {Hodaei}}, \bibinfo {author} {\bibfnamefont {M.-A.}\ \bibnamefont {Miri}},
  \bibinfo {author} {\bibfnamefont {M.}~\bibnamefont {Heinrich}}, \bibinfo
  {author} {\bibfnamefont {D.~N.}\ \bibnamefont {Christodoulides}},\ and\
  \bibinfo {author} {\bibfnamefont {M.}~\bibnamefont {Khajavikhan}},\
  }\bibfield  {title} {\bibinfo {title} {Parity-time--symmetric microring
  lasers},\ }\href@noop {} {\bibfield  {journal} {\bibinfo  {journal}
  {Science}\ }\textbf {\bibinfo {volume} {346}},\ \bibinfo {pages} {975}
  (\bibinfo {year} {2014})}\BibitemShut {NoStop}%
\bibitem [{\citenamefont {Feng}\ \emph {et~al.}(2014)\citenamefont {Feng},
  \citenamefont {Wong}, \citenamefont {Ma}, \citenamefont {Wang},\ and\
  \citenamefont {Zhang}}]{ar507}%
  \BibitemOpen
  \bibfield  {author} {\bibinfo {author} {\bibfnamefont {L.}~\bibnamefont
  {Feng}}, \bibinfo {author} {\bibfnamefont {Z.~J.}\ \bibnamefont {Wong}},
  \bibinfo {author} {\bibfnamefont {R.-M.}\ \bibnamefont {Ma}}, \bibinfo
  {author} {\bibfnamefont {Y.}~\bibnamefont {Wang}},\ and\ \bibinfo {author}
  {\bibfnamefont {X.}~\bibnamefont {Zhang}},\ }\bibfield  {title} {\bibinfo
  {title} {Single-mode laser by parity-time symmetry breaking},\ }\href@noop {}
  {\bibfield  {journal} {\bibinfo  {journal} {Science}\ }\textbf {\bibinfo
  {volume} {346}},\ \bibinfo {pages} {972} (\bibinfo {year}
  {2014})}\BibitemShut {NoStop}%
\bibitem [{\citenamefont {McDonald}\ and\ \citenamefont {Clerk}(2020)}]{ar301}%
  \BibitemOpen
  \bibfield  {author} {\bibinfo {author} {\bibfnamefont {A.}~\bibnamefont
  {McDonald}}\ and\ \bibinfo {author} {\bibfnamefont {A.~A.}\ \bibnamefont
  {Clerk}},\ }\bibfield  {title} {\bibinfo {title} {Exponentially-enhanced
  quantum sensing with non-hermitian lattice dynamics},\ }\href@noop {}
  {\bibfield  {journal} {\bibinfo  {journal} {Nature Communications}\ }\textbf
  {\bibinfo {volume} {11}},\ \bibinfo {pages} {5382} (\bibinfo {year}
  {2020})}\BibitemShut {NoStop}%
\bibitem [{\citenamefont {Ding}\ \emph {et~al.}(2021)\citenamefont {Ding},
  \citenamefont {Shi}, \citenamefont {Zhang}, \citenamefont {Shen},
  \citenamefont {Zhang},\ and\ \citenamefont {Zhang}}]{ar509}%
  \BibitemOpen
  \bibfield  {author} {\bibinfo {author} {\bibfnamefont {L.}~\bibnamefont
  {Ding}}, \bibinfo {author} {\bibfnamefont {K.}~\bibnamefont {Shi}}, \bibinfo
  {author} {\bibfnamefont {Q.}~\bibnamefont {Zhang}}, \bibinfo {author}
  {\bibfnamefont {D.}~\bibnamefont {Shen}}, \bibinfo {author} {\bibfnamefont
  {X.}~\bibnamefont {Zhang}},\ and\ \bibinfo {author} {\bibfnamefont
  {W.}~\bibnamefont {Zhang}},\ }\bibfield  {title} {\bibinfo {title}
  {Experimental determination of pt-symmetric exceptional points in a single
  trapped ion},\ }\href@noop {} {\bibfield  {journal} {\bibinfo  {journal}
  {Physical Review Letters}\ }\textbf {\bibinfo {volume} {126}},\ \bibinfo
  {pages} {083604} (\bibinfo {year} {2021})}\BibitemShut {NoStop}%
\bibitem [{\citenamefont {Wang}\ \emph {et~al.}(2021)\citenamefont {Wang},
  \citenamefont {Zhou}, \citenamefont {Zhang}, \citenamefont {Zhang},
  \citenamefont {Zhang}, \citenamefont {Xie}, \citenamefont {Wu}, \citenamefont
  {Chen}, \citenamefont {Ou}, \citenamefont {Wu} \emph {et~al.}}]{ar510}%
  \BibitemOpen
  \bibfield  {author} {\bibinfo {author} {\bibfnamefont {W.-C.}\ \bibnamefont
  {Wang}}, \bibinfo {author} {\bibfnamefont {Y.-L.}\ \bibnamefont {Zhou}},
  \bibinfo {author} {\bibfnamefont {H.-L.}\ \bibnamefont {Zhang}}, \bibinfo
  {author} {\bibfnamefont {J.}~\bibnamefont {Zhang}}, \bibinfo {author}
  {\bibfnamefont {M.-C.}\ \bibnamefont {Zhang}}, \bibinfo {author}
  {\bibfnamefont {Y.}~\bibnamefont {Xie}}, \bibinfo {author} {\bibfnamefont
  {C.-W.}\ \bibnamefont {Wu}}, \bibinfo {author} {\bibfnamefont
  {T.}~\bibnamefont {Chen}}, \bibinfo {author} {\bibfnamefont {B.-Q.}\
  \bibnamefont {Ou}}, \bibinfo {author} {\bibfnamefont {W.}~\bibnamefont {Wu}},
  \emph {et~al.},\ }\bibfield  {title} {\bibinfo {title} {Observation of
  pt-symmetric quantum coherence in a single-ion system},\ }\href@noop {}
  {\bibfield  {journal} {\bibinfo  {journal} {Physical Review A}\ }\textbf
  {\bibinfo {volume} {103}},\ \bibinfo {pages} {L020201} (\bibinfo {year}
  {2021})}\BibitemShut {NoStop}%
\bibitem [{\citenamefont {Liang}\ \emph {et~al.}(2023)\citenamefont {Liang},
  \citenamefont {Tang}, \citenamefont {Xu},\ and\ \citenamefont {Liu}}]{ar511}%
  \BibitemOpen
  \bibfield  {author} {\bibinfo {author} {\bibfnamefont {C.}~\bibnamefont
  {Liang}}, \bibinfo {author} {\bibfnamefont {Y.}~\bibnamefont {Tang}},
  \bibinfo {author} {\bibfnamefont {A.-N.}\ \bibnamefont {Xu}},\ and\ \bibinfo
  {author} {\bibfnamefont {Y.-C.}\ \bibnamefont {Liu}},\ }\bibfield  {title}
  {\bibinfo {title} {Observation of exceptional points in thermal atomic
  ensembles},\ }\href@noop {} {\bibfield  {journal} {\bibinfo  {journal}
  {Physical Review Letters}\ }\textbf {\bibinfo {volume} {130}},\ \bibinfo
  {pages} {263601} (\bibinfo {year} {2023})}\BibitemShut {NoStop}%
\bibitem [{\citenamefont {Li}\ \emph {et~al.}(2019)\citenamefont {Li},
  \citenamefont {Harter}, \citenamefont {Liu}, \citenamefont {de~Melo},
  \citenamefont {Joglekar},\ and\ \citenamefont {Luo}}]{ar512}%
  \BibitemOpen
  \bibfield  {author} {\bibinfo {author} {\bibfnamefont {J.}~\bibnamefont
  {Li}}, \bibinfo {author} {\bibfnamefont {A.~K.}\ \bibnamefont {Harter}},
  \bibinfo {author} {\bibfnamefont {J.}~\bibnamefont {Liu}}, \bibinfo {author}
  {\bibfnamefont {L.}~\bibnamefont {de~Melo}}, \bibinfo {author} {\bibfnamefont
  {Y.~N.}\ \bibnamefont {Joglekar}},\ and\ \bibinfo {author} {\bibfnamefont
  {L.}~\bibnamefont {Luo}},\ }\bibfield  {title} {\bibinfo {title} {Observation
  of parity-time symmetry breaking transitions in a dissipative floquet system
  of ultracold atoms},\ }\href@noop {} {\bibfield  {journal} {\bibinfo
  {journal} {Nature Communications}\ }\textbf {\bibinfo {volume} {10}},\
  \bibinfo {pages} {855} (\bibinfo {year} {2019})}\BibitemShut {NoStop}%
\bibitem [{\citenamefont {Zuo}\ \emph {et~al.}(2022)\citenamefont {Zuo},
  \citenamefont {Huang}, \citenamefont {Kuang}, \citenamefont {Xu},\ and\
  \citenamefont {Jing}}]{ar513}%
  \BibitemOpen
  \bibfield  {author} {\bibinfo {author} {\bibfnamefont {Y.}~\bibnamefont
  {Zuo}}, \bibinfo {author} {\bibfnamefont {R.}~\bibnamefont {Huang}}, \bibinfo
  {author} {\bibfnamefont {L.-M.}\ \bibnamefont {Kuang}}, \bibinfo {author}
  {\bibfnamefont {X.-W.}\ \bibnamefont {Xu}},\ and\ \bibinfo {author}
  {\bibfnamefont {H.}~\bibnamefont {Jing}},\ }\bibfield  {title} {\bibinfo
  {title} {Loss-induced suppression, revival, and switch of photon blockade},\
  }\href@noop {} {\bibfield  {journal} {\bibinfo  {journal} {Physical Review
  A}\ }\textbf {\bibinfo {volume} {106}},\ \bibinfo {pages} {043715} (\bibinfo
  {year} {2022})}\BibitemShut {NoStop}%
\bibitem [{\citenamefont {Yuan}\ \emph {et~al.}(2023)\citenamefont {Yuan},
  \citenamefont {Chen}, \citenamefont {Han}, \citenamefont {Wu}, \citenamefont
  {Li}, \citenamefont {Xia}, \citenamefont {Jiang},\ and\ \citenamefont
  {Song}}]{ar514}%
  \BibitemOpen
  \bibfield  {author} {\bibinfo {author} {\bibfnamefont {Z.-H.}\ \bibnamefont
  {Yuan}}, \bibinfo {author} {\bibfnamefont {Y.-J.}\ \bibnamefont {Chen}},
  \bibinfo {author} {\bibfnamefont {J.-X.}\ \bibnamefont {Han}}, \bibinfo
  {author} {\bibfnamefont {J.-L.}\ \bibnamefont {Wu}}, \bibinfo {author}
  {\bibfnamefont {W.-Q.}\ \bibnamefont {Li}}, \bibinfo {author} {\bibfnamefont
  {Y.}~\bibnamefont {Xia}}, \bibinfo {author} {\bibfnamefont {Y.-Y.}\
  \bibnamefont {Jiang}},\ and\ \bibinfo {author} {\bibfnamefont
  {J.}~\bibnamefont {Song}},\ }\bibfield  {title} {\bibinfo {title} {Periodic
  photon-magnon blockade in an optomagnonic system with chiral exceptional
  points},\ }\href@noop {} {\bibfield  {journal} {\bibinfo  {journal} {Physical
  Review B}\ }\textbf {\bibinfo {volume} {108}},\ \bibinfo {pages} {134409}
  (\bibinfo {year} {2023})}\BibitemShut {NoStop}%
\bibitem [{\citenamefont {Liu}\ \emph {et~al.}(2021)\citenamefont {Liu},
  \citenamefont {Wu}, \citenamefont {Duan}, \citenamefont {Rong},\ and\
  \citenamefont {Du}}]{ar516}%
  \BibitemOpen
  \bibfield  {author} {\bibinfo {author} {\bibfnamefont {W.}~\bibnamefont
  {Liu}}, \bibinfo {author} {\bibfnamefont {Y.}~\bibnamefont {Wu}}, \bibinfo
  {author} {\bibfnamefont {C.-K.}\ \bibnamefont {Duan}}, \bibinfo {author}
  {\bibfnamefont {X.}~\bibnamefont {Rong}},\ and\ \bibinfo {author}
  {\bibfnamefont {J.}~\bibnamefont {Du}},\ }\bibfield  {title} {\bibinfo
  {title} {Dynamically encircling an exceptional point in a real quantum
  system},\ }\href@noop {} {\bibfield  {journal} {\bibinfo  {journal} {Physical
  Review Letters}\ }\textbf {\bibinfo {volume} {126}},\ \bibinfo {pages}
  {170506} (\bibinfo {year} {2021})}\BibitemShut {NoStop}%
\bibitem [{\citenamefont {Xu}\ \emph {et~al.}(2016)\citenamefont {Xu},
  \citenamefont {Mason}, \citenamefont {Jiang},\ and\ \citenamefont
  {Harris}}]{ar517}%
  \BibitemOpen
  \bibfield  {author} {\bibinfo {author} {\bibfnamefont {H.}~\bibnamefont
  {Xu}}, \bibinfo {author} {\bibfnamefont {D.}~\bibnamefont {Mason}}, \bibinfo
  {author} {\bibfnamefont {L.}~\bibnamefont {Jiang}},\ and\ \bibinfo {author}
  {\bibfnamefont {J.}~\bibnamefont {Harris}},\ }\bibfield  {title} {\bibinfo
  {title} {Topological energy transfer in an optomechanical system with
  exceptional points},\ }\href@noop {} {\bibfield  {journal} {\bibinfo
  {journal} {Nature}\ }\textbf {\bibinfo {volume} {537}},\ \bibinfo {pages}
  {80} (\bibinfo {year} {2016})}\BibitemShut {NoStop}%
\bibitem [{\citenamefont {Bu}\ \emph {et~al.}(2024)\citenamefont {Bu},
  \citenamefont {Zhang}, \citenamefont {Ding}, \citenamefont {Li},
  \citenamefont {Zhang}, \citenamefont {Wang}, \citenamefont {Ding},
  \citenamefont {Yuan}, \citenamefont {Chen}, \citenamefont {Zhong} \emph
  {et~al.}}]{ar518}%
  \BibitemOpen
  \bibfield  {author} {\bibinfo {author} {\bibfnamefont {J.-T.}\ \bibnamefont
  {Bu}}, \bibinfo {author} {\bibfnamefont {J.-Q.}\ \bibnamefont {Zhang}},
  \bibinfo {author} {\bibfnamefont {G.-Y.}\ \bibnamefont {Ding}}, \bibinfo
  {author} {\bibfnamefont {J.-C.}\ \bibnamefont {Li}}, \bibinfo {author}
  {\bibfnamefont {J.-W.}\ \bibnamefont {Zhang}}, \bibinfo {author}
  {\bibfnamefont {B.}~\bibnamefont {Wang}}, \bibinfo {author} {\bibfnamefont
  {W.-Q.}\ \bibnamefont {Ding}}, \bibinfo {author} {\bibfnamefont {W.-F.}\
  \bibnamefont {Yuan}}, \bibinfo {author} {\bibfnamefont {L.}~\bibnamefont
  {Chen}}, \bibinfo {author} {\bibfnamefont {Q.}~\bibnamefont {Zhong}}, \emph
  {et~al.},\ }\bibfield  {title} {\bibinfo {title} {Chiral quantum heating and
  cooling with an optically controlled ion},\ }\href@noop {} {\bibfield
  {journal} {\bibinfo  {journal} {Light: Science \& Applications}\ }\textbf
  {\bibinfo {volume} {13}},\ \bibinfo {pages} {143} (\bibinfo {year}
  {2024})}\BibitemShut {NoStop}%
\bibitem [{\citenamefont {Khandelwal}\ \emph {et~al.}(2021)\citenamefont
  {Khandelwal}, \citenamefont {Brunner},\ and\ \citenamefont {Haack}}]{ar519}%
  \BibitemOpen
  \bibfield  {author} {\bibinfo {author} {\bibfnamefont {S.}~\bibnamefont
  {Khandelwal}}, \bibinfo {author} {\bibfnamefont {N.}~\bibnamefont
  {Brunner}},\ and\ \bibinfo {author} {\bibfnamefont {G.}~\bibnamefont
  {Haack}},\ }\bibfield  {title} {\bibinfo {title} {Signatures of liouvillian
  exceptional points in a quantum thermal machine},\ }\href@noop {} {\bibfield
  {journal} {\bibinfo  {journal} {PRX Quantum}\ }\textbf {\bibinfo {volume}
  {2}},\ \bibinfo {pages} {040346} (\bibinfo {year} {2021})}\BibitemShut
  {NoStop}%
\bibitem [{\citenamefont {Ju}\ \emph {et~al.}(2019)\citenamefont {Ju},
  \citenamefont {Miranowicz}, \citenamefont {Chen},\ and\ \citenamefont
  {Nori}}]{JuPRA2019}%
  \BibitemOpen
  \bibfield  {author} {\bibinfo {author} {\bibfnamefont {C.-Y.}\ \bibnamefont
  {Ju}}, \bibinfo {author} {\bibfnamefont {A.}~\bibnamefont {Miranowicz}},
  \bibinfo {author} {\bibfnamefont {G.-Y.}\ \bibnamefont {Chen}},\ and\
  \bibinfo {author} {\bibfnamefont {F.}~\bibnamefont {Nori}},\ }\bibfield
  {title} {\bibinfo {title} {Non-hermitian hamiltonians and no-go theorems in
  quantum information},\ }\href {https://doi.org/10.1103/PhysRevA.100.062118}
  {\bibfield  {journal} {\bibinfo  {journal} {Phys. Rev. A}\ }\textbf {\bibinfo
  {volume} {100}},\ \bibinfo {pages} {062118} (\bibinfo {year}
  {2019})}\BibitemShut {NoStop}%
\bibitem [{\citenamefont {Chakraborty}\ and\ \citenamefont
  {Sarma}(2019)}]{ar508}%
  \BibitemOpen
  \bibfield  {author} {\bibinfo {author} {\bibfnamefont {S.}~\bibnamefont
  {Chakraborty}}\ and\ \bibinfo {author} {\bibfnamefont {A.~K.}\ \bibnamefont
  {Sarma}},\ }\bibfield  {title} {\bibinfo {title} {Delayed sudden death of
  entanglement at exceptional points},\ }\href@noop {} {\bibfield  {journal}
  {\bibinfo  {journal} {Physical Review A}\ }\textbf {\bibinfo {volume}
  {100}},\ \bibinfo {pages} {063846} (\bibinfo {year} {2019})}\BibitemShut
  {NoStop}%
\bibitem [{\citenamefont {Zhang}\ \emph {et~al.}(2024)\citenamefont {Zhang},
  \citenamefont {Zhang}, \citenamefont {Wei}, \citenamefont {Liang},
  \citenamefont {Mei},\ and\ \citenamefont {Yang}}]{ar04}%
  \BibitemOpen
  \bibfield  {author} {\bibinfo {author} {\bibfnamefont {Y.-X.}\ \bibnamefont
  {Zhang}}, \bibinfo {author} {\bibfnamefont {Z.-T.}\ \bibnamefont {Zhang}},
  \bibinfo {author} {\bibfnamefont {X.-Z.}\ \bibnamefont {Wei}}, \bibinfo
  {author} {\bibfnamefont {B.-L.}\ \bibnamefont {Liang}}, \bibinfo {author}
  {\bibfnamefont {F.}~\bibnamefont {Mei}},\ and\ \bibinfo {author}
  {\bibfnamefont {Z.-S.}\ \bibnamefont {Yang}},\ }\bibfield  {title} {\bibinfo
  {title} {Entanglement dynamics of two non-hermitian qubits},\ }\href@noop {}
  {\bibfield  {journal} {\bibinfo  {journal} {Journal of Physics B: Atomic,
  Molecular and Optical Physics}\ }\textbf {\bibinfo {volume} {57}},\ \bibinfo
  {pages} {085501} (\bibinfo {year} {2024})}\BibitemShut {NoStop}%
\bibitem [{\citenamefont {Kumar}\ \emph {et~al.}(2022)\citenamefont {Kumar},
  \citenamefont {Murch},\ and\ \citenamefont {Joglekar}}]{ar6}%
  \BibitemOpen
  \bibfield  {author} {\bibinfo {author} {\bibfnamefont {A.}~\bibnamefont
  {Kumar}}, \bibinfo {author} {\bibfnamefont {K.~W.}\ \bibnamefont {Murch}},\
  and\ \bibinfo {author} {\bibfnamefont {Y.~N.}\ \bibnamefont {Joglekar}},\
  }\bibfield  {title} {\bibinfo {title} {Maximal quantum entanglement at
  exceptional points via unitary and thermal dynamics},\ }\href@noop {}
  {\bibfield  {journal} {\bibinfo  {journal} {Physical Review A}\ }\textbf
  {\bibinfo {volume} {105}},\ \bibinfo {pages} {012422} (\bibinfo {year}
  {2022})}\BibitemShut {NoStop}%
\bibitem [{\citenamefont {Liu}\ \emph {et~al.}(2024)\citenamefont {Liu},
  \citenamefont {Su}, \citenamefont {Zuo}, \citenamefont {Chen}, \citenamefont
  {{\"O}zdemir},\ and\ \citenamefont {Jing}}]{ar26}%
  \BibitemOpen
  \bibfield  {author} {\bibinfo {author} {\bibfnamefont {B.-B.}\ \bibnamefont
  {Liu}}, \bibinfo {author} {\bibfnamefont {S.-L.}\ \bibnamefont {Su}},
  \bibinfo {author} {\bibfnamefont {Y.-L.}\ \bibnamefont {Zuo}}, \bibinfo
  {author} {\bibfnamefont {G.}~\bibnamefont {Chen}}, \bibinfo {author}
  {\bibfnamefont {{\c{S}}.}~\bibnamefont {{\"O}zdemir}},\ and\ \bibinfo
  {author} {\bibfnamefont {H.}~\bibnamefont {Jing}},\ }\bibfield  {title}
  {\bibinfo {title} {Faster preparation of multi-qubit entanglement with higher
  success rates},\ }\href@noop {} {\bibfield  {journal} {\bibinfo  {journal}
  {arXiv preprint arXiv:2407.08525}\ } (\bibinfo {year} {2024})}\BibitemShut
  {NoStop}%
\bibitem [{\citenamefont {Khandelwal}\ \emph {et~al.}(2024)\citenamefont
  {Khandelwal}, \citenamefont {Chen}, \citenamefont {Murch},\ and\
  \citenamefont {Haack}}]{ar41}%
  \BibitemOpen
  \bibfield  {author} {\bibinfo {author} {\bibfnamefont {S.}~\bibnamefont
  {Khandelwal}}, \bibinfo {author} {\bibfnamefont {W.}~\bibnamefont {Chen}},
  \bibinfo {author} {\bibfnamefont {K.~W.}\ \bibnamefont {Murch}},\ and\
  \bibinfo {author} {\bibfnamefont {G.}~\bibnamefont {Haack}},\ }\bibfield
  {title} {\bibinfo {title} {Chiral bell-state transfer via dissipative
  liouvillian dynamics},\ }\href@noop {} {\bibfield  {journal} {\bibinfo
  {journal} {Physical Review Letters}\ }\textbf {\bibinfo {volume} {133}},\
  \bibinfo {pages} {070403} (\bibinfo {year} {2024})}\BibitemShut {NoStop}%
\bibitem [{\citenamefont {Zhao}\ \emph {et~al.}(2021)\citenamefont {Zhao},
  \citenamefont {Zhang}, \citenamefont {Chen}, \citenamefont {Wang},\ and\
  \citenamefont {Hu}}]{ar520}%
  \BibitemOpen
  \bibfield  {author} {\bibinfo {author} {\bibfnamefont {Y.}~\bibnamefont
  {Zhao}}, \bibinfo {author} {\bibfnamefont {R.}~\bibnamefont {Zhang}},
  \bibinfo {author} {\bibfnamefont {W.}~\bibnamefont {Chen}}, \bibinfo {author}
  {\bibfnamefont {X.-B.}\ \bibnamefont {Wang}},\ and\ \bibinfo {author}
  {\bibfnamefont {J.}~\bibnamefont {Hu}},\ }\bibfield  {title} {\bibinfo
  {title} {Creation of greenberger-horne-zeilinger states with thousands of
  atoms by entanglement amplification},\ }\href@noop {} {\bibfield  {journal}
  {\bibinfo  {journal} {NPJ Quantum Information}\ }\textbf {\bibinfo {volume}
  {7}},\ \bibinfo {pages} {24} (\bibinfo {year} {2021})}\BibitemShut {NoStop}%
\bibitem [{\citenamefont {Friis}\ \emph {et~al.}(2019)\citenamefont {Friis},
  \citenamefont {Vitagliano}, \citenamefont {Malik},\ and\ \citenamefont
  {Huber}}]{ar521}%
  \BibitemOpen
  \bibfield  {author} {\bibinfo {author} {\bibfnamefont {N.}~\bibnamefont
  {Friis}}, \bibinfo {author} {\bibfnamefont {G.}~\bibnamefont {Vitagliano}},
  \bibinfo {author} {\bibfnamefont {M.}~\bibnamefont {Malik}},\ and\ \bibinfo
  {author} {\bibfnamefont {M.}~\bibnamefont {Huber}},\ }\bibfield  {title}
  {\bibinfo {title} {Entanglement certification from theory to experiment},\
  }\href@noop {} {\bibfield  {journal} {\bibinfo  {journal} {Nature Reviews
  Physics}\ }\textbf {\bibinfo {volume} {1}},\ \bibinfo {pages} {72} (\bibinfo
  {year} {2019})}\BibitemShut {NoStop}%
\bibitem [{\citenamefont {McCutcheon}\ \emph {et~al.}(2016)\citenamefont
  {McCutcheon}, \citenamefont {Pappa}, \citenamefont {Bell}, \citenamefont
  {Mcmillan}, \citenamefont {Chailloux}, \citenamefont {Lawson}, \citenamefont
  {Mafu}, \citenamefont {Markham}, \citenamefont {Diamanti}, \citenamefont
  {Kerenidis} \emph {et~al.}}]{ar522}%
  \BibitemOpen
  \bibfield  {author} {\bibinfo {author} {\bibfnamefont {W.}~\bibnamefont
  {McCutcheon}}, \bibinfo {author} {\bibfnamefont {A.}~\bibnamefont {Pappa}},
  \bibinfo {author} {\bibfnamefont {B.~A.}\ \bibnamefont {Bell}}, \bibinfo
  {author} {\bibfnamefont {A.}~\bibnamefont {Mcmillan}}, \bibinfo {author}
  {\bibfnamefont {A.}~\bibnamefont {Chailloux}}, \bibinfo {author}
  {\bibfnamefont {T.}~\bibnamefont {Lawson}}, \bibinfo {author} {\bibfnamefont
  {M.}~\bibnamefont {Mafu}}, \bibinfo {author} {\bibfnamefont {D.}~\bibnamefont
  {Markham}}, \bibinfo {author} {\bibfnamefont {E.}~\bibnamefont {Diamanti}},
  \bibinfo {author} {\bibfnamefont {I.}~\bibnamefont {Kerenidis}}, \emph
  {et~al.},\ }\bibfield  {title} {\bibinfo {title} {Experimental verification
  of multipartite entanglement in quantum networks},\ }\href@noop {} {\bibfield
   {journal} {\bibinfo  {journal} {Nature Communications}\ }\textbf {\bibinfo
  {volume} {7}},\ \bibinfo {pages} {13251} (\bibinfo {year}
  {2016})}\BibitemShut {NoStop}%
\bibitem [{\citenamefont {Cervera-Lierta}\ \emph {et~al.}(2022)\citenamefont
  {Cervera-Lierta}, \citenamefont {Krenn}, \citenamefont {Aspuru-Guzik},\ and\
  \citenamefont {Galda}}]{ar523}%
  \BibitemOpen
  \bibfield  {author} {\bibinfo {author} {\bibfnamefont {A.}~\bibnamefont
  {Cervera-Lierta}}, \bibinfo {author} {\bibfnamefont {M.}~\bibnamefont
  {Krenn}}, \bibinfo {author} {\bibfnamefont {A.}~\bibnamefont
  {Aspuru-Guzik}},\ and\ \bibinfo {author} {\bibfnamefont {A.}~\bibnamefont
  {Galda}},\ }\bibfield  {title} {\bibinfo {title} {Experimental
  high-dimensional greenberger-horne-zeilinger entanglement with
  superconducting transmon qutrits},\ }\href@noop {} {\bibfield  {journal}
  {\bibinfo  {journal} {Physical Review Applied}\ }\textbf {\bibinfo {volume}
  {17}},\ \bibinfo {pages} {024062} (\bibinfo {year} {2022})}\BibitemShut
  {NoStop}%
\bibitem [{\citenamefont {Zhao}\ \emph {et~al.}(2024)\citenamefont {Zhao},
  \citenamefont {Yang}, \citenamefont {Li},\ and\ \citenamefont
  {Shao}}]{ar526}%
  \BibitemOpen
  \bibfield  {author} {\bibinfo {author} {\bibfnamefont {Y.}~\bibnamefont
  {Zhao}}, \bibinfo {author} {\bibfnamefont {Y.-Q.}\ \bibnamefont {Yang}},
  \bibinfo {author} {\bibfnamefont {W.}~\bibnamefont {Li}},\ and\ \bibinfo
  {author} {\bibfnamefont {X.-Q.}\ \bibnamefont {Shao}},\ }\bibfield  {title}
  {\bibinfo {title} {Dissipative stabilization of high-dimensional ghz states
  for neutral atoms},\ }\href@noop {} {\bibfield  {journal} {\bibinfo
  {journal} {Applied Physics Letters}\ }\textbf {\bibinfo {volume} {124}}
  (\bibinfo {year} {2024})}\BibitemShut {NoStop}%
\bibitem [{\citenamefont {Ku}\ \emph {et~al.}(2020)\citenamefont {Ku},
  \citenamefont {Lambert}, \citenamefont {Chan}, \citenamefont {Emary},
  \citenamefont {Chen},\ and\ \citenamefont {Nori}}]{ar36}%
  \BibitemOpen
  \bibfield  {author} {\bibinfo {author} {\bibfnamefont {H.-Y.}\ \bibnamefont
  {Ku}}, \bibinfo {author} {\bibfnamefont {N.}~\bibnamefont {Lambert}},
  \bibinfo {author} {\bibfnamefont {F.-J.}\ \bibnamefont {Chan}}, \bibinfo
  {author} {\bibfnamefont {C.}~\bibnamefont {Emary}}, \bibinfo {author}
  {\bibfnamefont {Y.-N.}\ \bibnamefont {Chen}},\ and\ \bibinfo {author}
  {\bibfnamefont {F.}~\bibnamefont {Nori}},\ }\bibfield  {title} {\bibinfo
  {title} {Experimental test of non-macrorealistic cat states in the cloud},\
  }\href@noop {} {\bibfield  {journal} {\bibinfo  {journal} {npj Quantum
  information}\ }\textbf {\bibinfo {volume} {6}},\ \bibinfo {pages} {98}
  (\bibinfo {year} {2020})}\BibitemShut {NoStop}%
\bibitem [{\citenamefont {Yang}\ \emph {et~al.}(2022)\citenamefont {Yang},
  \citenamefont {Ku}, \citenamefont {Baishya}, \citenamefont {Zhang},
  \citenamefont {Kockum}, \citenamefont {Chen}, \citenamefont {Li},
  \citenamefont {Tsai},\ and\ \citenamefont {Nori}}]{ar37}%
  \BibitemOpen
  \bibfield  {author} {\bibinfo {author} {\bibfnamefont {Z.-P.}\ \bibnamefont
  {Yang}}, \bibinfo {author} {\bibfnamefont {H.-Y.}\ \bibnamefont {Ku}},
  \bibinfo {author} {\bibfnamefont {A.}~\bibnamefont {Baishya}}, \bibinfo
  {author} {\bibfnamefont {Y.-R.}\ \bibnamefont {Zhang}}, \bibinfo {author}
  {\bibfnamefont {A.~F.}\ \bibnamefont {Kockum}}, \bibinfo {author}
  {\bibfnamefont {Y.-N.}\ \bibnamefont {Chen}}, \bibinfo {author}
  {\bibfnamefont {F.-L.}\ \bibnamefont {Li}}, \bibinfo {author} {\bibfnamefont
  {J.-S.}\ \bibnamefont {Tsai}},\ and\ \bibinfo {author} {\bibfnamefont
  {F.}~\bibnamefont {Nori}},\ }\bibfield  {title} {\bibinfo {title}
  {Deterministic one-way logic gates on a cloud quantum computer},\ }\href@noop
  {} {\bibfield  {journal} {\bibinfo  {journal} {Physical Review A}\ }\textbf
  {\bibinfo {volume} {105}},\ \bibinfo {pages} {042610} (\bibinfo {year}
  {2022})}\BibitemShut {NoStop}%
\bibitem [{\citenamefont {Yamasaki}\ \emph {et~al.}(2018)\citenamefont
  {Yamasaki}, \citenamefont {Pirker}, \citenamefont {Murao}, \citenamefont
  {D{\"u}r},\ and\ \citenamefont {Kraus}}]{ar524}%
  \BibitemOpen
  \bibfield  {author} {\bibinfo {author} {\bibfnamefont {H.}~\bibnamefont
  {Yamasaki}}, \bibinfo {author} {\bibfnamefont {A.}~\bibnamefont {Pirker}},
  \bibinfo {author} {\bibfnamefont {M.}~\bibnamefont {Murao}}, \bibinfo
  {author} {\bibfnamefont {W.}~\bibnamefont {D{\"u}r}},\ and\ \bibinfo {author}
  {\bibfnamefont {B.}~\bibnamefont {Kraus}},\ }\bibfield  {title} {\bibinfo
  {title} {Multipartite entanglement outperforming bipartite entanglement under
  limited quantum system sizes},\ }\href@noop {} {\bibfield  {journal}
  {\bibinfo  {journal} {Physical Review A}\ }\textbf {\bibinfo {volume} {98}},\
  \bibinfo {pages} {052313} (\bibinfo {year} {2018})}\BibitemShut {NoStop}%
\bibitem [{\citenamefont {Chen}\ \emph
  {et~al.}(2005{\natexlab{a}})\citenamefont {Chen}, \citenamefont {Zhang},
  \citenamefont {Zhao}, \citenamefont {Zhou}, \citenamefont {Lu}, \citenamefont
  {Peng}, \citenamefont {Yang},\ and\ \citenamefont {Pan}}]{ar525}%
  \BibitemOpen
  \bibfield  {author} {\bibinfo {author} {\bibfnamefont {Y.-A.}\ \bibnamefont
  {Chen}}, \bibinfo {author} {\bibfnamefont {A.-N.}\ \bibnamefont {Zhang}},
  \bibinfo {author} {\bibfnamefont {Z.}~\bibnamefont {Zhao}}, \bibinfo {author}
  {\bibfnamefont {X.-Q.}\ \bibnamefont {Zhou}}, \bibinfo {author}
  {\bibfnamefont {C.-Y.}\ \bibnamefont {Lu}}, \bibinfo {author} {\bibfnamefont
  {C.-Z.}\ \bibnamefont {Peng}}, \bibinfo {author} {\bibfnamefont
  {T.}~\bibnamefont {Yang}},\ and\ \bibinfo {author} {\bibfnamefont {J.-W.}\
  \bibnamefont {Pan}},\ }\bibfield  {title} {\bibinfo {title} {Experimental
  quantum secret sharing and third-man quantum cryptography},\ }\href@noop {}
  {\bibfield  {journal} {\bibinfo  {journal} {Physical Review Letters}\
  }\textbf {\bibinfo {volume} {95}},\ \bibinfo {pages} {200502} (\bibinfo
  {year} {2005}{\natexlab{a}})}\BibitemShut {NoStop}%
\bibitem [{\citenamefont {Miguel-Ramiro}\ \emph {et~al.}(2021)\citenamefont
  {Miguel-Ramiro}, \citenamefont {Pirker},\ and\ \citenamefont
  {D{\"u}r}}]{ar29}%
  \BibitemOpen
  \bibfield  {author} {\bibinfo {author} {\bibfnamefont {J.}~\bibnamefont
  {Miguel-Ramiro}}, \bibinfo {author} {\bibfnamefont {A.}~\bibnamefont
  {Pirker}},\ and\ \bibinfo {author} {\bibfnamefont {W.}~\bibnamefont
  {D{\"u}r}},\ }\bibfield  {title} {\bibinfo {title} {Genuine quantum networks
  with superposed tasks and addressing},\ }\href@noop {} {\bibfield  {journal}
  {\bibinfo  {journal} {npj Quantum Information}\ }\textbf {\bibinfo {volume}
  {7}},\ \bibinfo {pages} {135} (\bibinfo {year} {2021})}\BibitemShut {NoStop}%
\bibitem [{\citenamefont {Kao}\ \emph {et~al.}(2024)\citenamefont {Kao},
  \citenamefont {Huang}, \citenamefont {Tsai}, \citenamefont {Chen},
  \citenamefont {Sun}, \citenamefont {Li}, \citenamefont {Liao}, \citenamefont
  {Chuu}, \citenamefont {Lu},\ and\ \citenamefont {Li}}]{ar32}%
  \BibitemOpen
  \bibfield  {author} {\bibinfo {author} {\bibfnamefont {W.-T.}\ \bibnamefont
  {Kao}}, \bibinfo {author} {\bibfnamefont {C.-Y.}\ \bibnamefont {Huang}},
  \bibinfo {author} {\bibfnamefont {T.-J.}\ \bibnamefont {Tsai}}, \bibinfo
  {author} {\bibfnamefont {S.-H.}\ \bibnamefont {Chen}}, \bibinfo {author}
  {\bibfnamefont {S.-Y.}\ \bibnamefont {Sun}}, \bibinfo {author} {\bibfnamefont
  {Y.-C.}\ \bibnamefont {Li}}, \bibinfo {author} {\bibfnamefont {T.-L.}\
  \bibnamefont {Liao}}, \bibinfo {author} {\bibfnamefont {C.-S.}\ \bibnamefont
  {Chuu}}, \bibinfo {author} {\bibfnamefont {H.}~\bibnamefont {Lu}},\ and\
  \bibinfo {author} {\bibfnamefont {C.-M.}\ \bibnamefont {Li}},\ }\bibfield
  {title} {\bibinfo {title} {Scalable determination of multipartite
  entanglement in quantum networks},\ }\href@noop {} {\bibfield  {journal}
  {\bibinfo  {journal} {npj Quantum Information}\ }\textbf {\bibinfo {volume}
  {10}},\ \bibinfo {pages} {73} (\bibinfo {year} {2024})}\BibitemShut {NoStop}%
\bibitem [{\citenamefont {Chen}\ \emph
  {et~al.}(2005{\natexlab{b}})\citenamefont {Chen}, \citenamefont {Zhang},
  \citenamefont {Zhao}, \citenamefont {Zhou}, \citenamefont {Lu}, \citenamefont
  {Peng}, \citenamefont {Yang},\ and\ \citenamefont {Pan}}]{ar30}%
  \BibitemOpen
  \bibfield  {author} {\bibinfo {author} {\bibfnamefont {Y.-A.}\ \bibnamefont
  {Chen}}, \bibinfo {author} {\bibfnamefont {A.-N.}\ \bibnamefont {Zhang}},
  \bibinfo {author} {\bibfnamefont {Z.}~\bibnamefont {Zhao}}, \bibinfo {author}
  {\bibfnamefont {X.-Q.}\ \bibnamefont {Zhou}}, \bibinfo {author}
  {\bibfnamefont {C.-Y.}\ \bibnamefont {Lu}}, \bibinfo {author} {\bibfnamefont
  {C.-Z.}\ \bibnamefont {Peng}}, \bibinfo {author} {\bibfnamefont
  {T.}~\bibnamefont {Yang}},\ and\ \bibinfo {author} {\bibfnamefont {J.-W.}\
  \bibnamefont {Pan}},\ }\bibfield  {title} {\bibinfo {title} {Experimental
  quantum secret sharing and third-man quantum cryptography},\ }\href@noop {}
  {\bibfield  {journal} {\bibinfo  {journal} {Physical Leview Letters}\
  }\textbf {\bibinfo {volume} {95}},\ \bibinfo {pages} {200502} (\bibinfo
  {year} {2005}{\natexlab{b}})}\BibitemShut {NoStop}%
\bibitem [{\citenamefont {Markham}\ and\ \citenamefont {Sanders}(2008)}]{ar31}%
  \BibitemOpen
  \bibfield  {author} {\bibinfo {author} {\bibfnamefont {D.}~\bibnamefont
  {Markham}}\ and\ \bibinfo {author} {\bibfnamefont {B.~C.}\ \bibnamefont
  {Sanders}},\ }\bibfield  {title} {\bibinfo {title} {Graph states for quantum
  secret sharing},\ }\href@noop {} {\bibfield  {journal} {\bibinfo  {journal}
  {Physical Review A}\ }\textbf {\bibinfo {volume} {78}},\ \bibinfo {pages}
  {042309} (\bibinfo {year} {2008})}\BibitemShut {NoStop}%
\bibitem [{\citenamefont {Lo}\ \emph {et~al.}(2014)\citenamefont {Lo},
  \citenamefont {Curty},\ and\ \citenamefont {Tamaki}}]{ar33}%
  \BibitemOpen
  \bibfield  {author} {\bibinfo {author} {\bibfnamefont {H.-K.}\ \bibnamefont
  {Lo}}, \bibinfo {author} {\bibfnamefont {M.}~\bibnamefont {Curty}},\ and\
  \bibinfo {author} {\bibfnamefont {K.}~\bibnamefont {Tamaki}},\ }\bibfield
  {title} {\bibinfo {title} {Secure quantum key distribution},\ }\href@noop {}
  {\bibfield  {journal} {\bibinfo  {journal} {Nature Photonics}\ }\textbf
  {\bibinfo {volume} {8}},\ \bibinfo {pages} {595} (\bibinfo {year}
  {2014})}\BibitemShut {NoStop}%
\bibitem [{\citenamefont {Epping}\ \emph {et~al.}(2017)\citenamefont {Epping},
  \citenamefont {Kampermann}, \citenamefont {Bru{\ss}} \emph {et~al.}}]{ar34}%
  \BibitemOpen
  \bibfield  {author} {\bibinfo {author} {\bibfnamefont {M.}~\bibnamefont
  {Epping}}, \bibinfo {author} {\bibfnamefont {H.}~\bibnamefont {Kampermann}},
  \bibinfo {author} {\bibfnamefont {D.}~\bibnamefont {Bru{\ss}}}, \emph
  {et~al.},\ }\bibfield  {title} {\bibinfo {title} {Multi-partite entanglement
  can speed up quantum key distribution in networks},\ }\href@noop {}
  {\bibfield  {journal} {\bibinfo  {journal} {New Journal of Physics}\ }\textbf
  {\bibinfo {volume} {19}},\ \bibinfo {pages} {093012} (\bibinfo {year}
  {2017})}\BibitemShut {NoStop}%
\bibitem [{\citenamefont {Ciampini}\ \emph {et~al.}(2017)\citenamefont
  {Ciampini}, \citenamefont {Mancino}, \citenamefont {Orieux}, \citenamefont
  {Vigliar}, \citenamefont {Mataloni}, \citenamefont {Paternostro},\ and\
  \citenamefont {Barbieri}}]{ar35}%
  \BibitemOpen
  \bibfield  {author} {\bibinfo {author} {\bibfnamefont {M.~A.}\ \bibnamefont
  {Ciampini}}, \bibinfo {author} {\bibfnamefont {L.}~\bibnamefont {Mancino}},
  \bibinfo {author} {\bibfnamefont {A.}~\bibnamefont {Orieux}}, \bibinfo
  {author} {\bibfnamefont {C.}~\bibnamefont {Vigliar}}, \bibinfo {author}
  {\bibfnamefont {P.}~\bibnamefont {Mataloni}}, \bibinfo {author}
  {\bibfnamefont {M.}~\bibnamefont {Paternostro}},\ and\ \bibinfo {author}
  {\bibfnamefont {M.}~\bibnamefont {Barbieri}},\ }\bibfield  {title} {\bibinfo
  {title} {Experimental extractable work-based multipartite separability
  criteria},\ }\href@noop {} {\bibfield  {journal} {\bibinfo  {journal} {npj
  Quantum Information}\ }\textbf {\bibinfo {volume} {3}},\ \bibinfo {pages}
  {10} (\bibinfo {year} {2017})}\BibitemShut {NoStop}%
\bibitem [{\citenamefont {Islam}\ \emph {et~al.}(2015)\citenamefont {Islam},
  \citenamefont {Ma}, \citenamefont {Preiss}, \citenamefont {Eric~Tai},
  \citenamefont {Lukin}, \citenamefont {Rispoli},\ and\ \citenamefont
  {Greiner}}]{ar7}%
  \BibitemOpen
  \bibfield  {author} {\bibinfo {author} {\bibfnamefont {R.}~\bibnamefont
  {Islam}}, \bibinfo {author} {\bibfnamefont {R.}~\bibnamefont {Ma}}, \bibinfo
  {author} {\bibfnamefont {P.~M.}\ \bibnamefont {Preiss}}, \bibinfo {author}
  {\bibfnamefont {M.}~\bibnamefont {Eric~Tai}}, \bibinfo {author}
  {\bibfnamefont {A.}~\bibnamefont {Lukin}}, \bibinfo {author} {\bibfnamefont
  {M.}~\bibnamefont {Rispoli}},\ and\ \bibinfo {author} {\bibfnamefont
  {M.}~\bibnamefont {Greiner}},\ }\bibfield  {title} {\bibinfo {title}
  {Measuring entanglement entropy in a quantum many-body system},\ }\href@noop
  {} {\bibfield  {journal} {\bibinfo  {journal} {Nature}\ }\textbf {\bibinfo
  {volume} {528}},\ \bibinfo {pages} {77} (\bibinfo {year} {2015})}\BibitemShut
  {NoStop}%
\bibitem [{\citenamefont {Coffman}\ \emph {et~al.}(2000)\citenamefont
  {Coffman}, \citenamefont {Kundu},\ and\ \citenamefont {Wootters}}]{ar10}%
  \BibitemOpen
  \bibfield  {author} {\bibinfo {author} {\bibfnamefont {V.}~\bibnamefont
  {Coffman}}, \bibinfo {author} {\bibfnamefont {J.}~\bibnamefont {Kundu}},\
  and\ \bibinfo {author} {\bibfnamefont {W.~K.}\ \bibnamefont {Wootters}},\
  }\bibfield  {title} {\bibinfo {title} {Distributed entanglement},\
  }\href@noop {} {\bibfield  {journal} {\bibinfo  {journal} {Physical Review
  A}\ }\textbf {\bibinfo {volume} {61}},\ \bibinfo {pages} {052306} (\bibinfo
  {year} {2000})}\BibitemShut {NoStop}%
\bibitem [{\citenamefont {D{\"u}r}\ \emph {et~al.}(2000)\citenamefont
  {D{\"u}r}, \citenamefont {Vidal},\ and\ \citenamefont {Cirac}}]{ar13}%
  \BibitemOpen
  \bibfield  {author} {\bibinfo {author} {\bibfnamefont {W.}~\bibnamefont
  {D{\"u}r}}, \bibinfo {author} {\bibfnamefont {G.}~\bibnamefont {Vidal}},\
  and\ \bibinfo {author} {\bibfnamefont {J.~I.}\ \bibnamefont {Cirac}},\
  }\bibfield  {title} {\bibinfo {title} {Three qubits can be entangled in two
  inequivalent ways},\ }\href@noop {} {\bibfield  {journal} {\bibinfo
  {journal} {Physical Review A}\ }\textbf {\bibinfo {volume} {62}},\ \bibinfo
  {pages} {062314} (\bibinfo {year} {2000})}\BibitemShut {NoStop}%
\bibitem [{\citenamefont {Kjaergaard}\ \emph {et~al.}(2020)\citenamefont
  {Kjaergaard}, \citenamefont {Schwartz}, \citenamefont {Braum{\"u}ller},
  \citenamefont {Krantz}, \citenamefont {Wang}, \citenamefont {Gustavsson},\
  and\ \citenamefont {Oliver}}]{ar4}%
  \BibitemOpen
  \bibfield  {author} {\bibinfo {author} {\bibfnamefont {M.}~\bibnamefont
  {Kjaergaard}}, \bibinfo {author} {\bibfnamefont {M.~E.}\ \bibnamefont
  {Schwartz}}, \bibinfo {author} {\bibfnamefont {J.}~\bibnamefont
  {Braum{\"u}ller}}, \bibinfo {author} {\bibfnamefont {P.}~\bibnamefont
  {Krantz}}, \bibinfo {author} {\bibfnamefont {J.~I.-J.}\ \bibnamefont {Wang}},
  \bibinfo {author} {\bibfnamefont {S.}~\bibnamefont {Gustavsson}},\ and\
  \bibinfo {author} {\bibfnamefont {W.~D.}\ \bibnamefont {Oliver}},\ }\bibfield
   {title} {\bibinfo {title} {Superconducting qubits: Current state of play},\
  }\href@noop {} {\bibfield  {journal} {\bibinfo  {journal} {Annual Review of
  Condensed Matter Physics}\ }\textbf {\bibinfo {volume} {11}},\ \bibinfo
  {pages} {369} (\bibinfo {year} {2020})}\BibitemShut {NoStop}%
\bibitem [{\citenamefont {Guo}\ \emph {et~al.}(2009)\citenamefont {Guo},
  \citenamefont {Salamo}, \citenamefont {Duchesne}, \citenamefont {Morandotti},
  \citenamefont {Volatier-Ravat}, \citenamefont {Aimez}, \citenamefont
  {Siviloglou},\ and\ \citenamefont {Christodoulides}}]{ar40}%
  \BibitemOpen
  \bibfield  {author} {\bibinfo {author} {\bibfnamefont {A.}~\bibnamefont
  {Guo}}, \bibinfo {author} {\bibfnamefont {G.~J.}\ \bibnamefont {Salamo}},
  \bibinfo {author} {\bibfnamefont {D.}~\bibnamefont {Duchesne}}, \bibinfo
  {author} {\bibfnamefont {R.}~\bibnamefont {Morandotti}}, \bibinfo {author}
  {\bibfnamefont {M.}~\bibnamefont {Volatier-Ravat}}, \bibinfo {author}
  {\bibfnamefont {V.}~\bibnamefont {Aimez}}, \bibinfo {author} {\bibfnamefont
  {G.~A.}\ \bibnamefont {Siviloglou}},\ and\ \bibinfo {author} {\bibfnamefont
  {D.~N.}\ \bibnamefont {Christodoulides}},\ }\bibfield  {title} {\bibinfo
  {title} {Observation of pt-symmetry breaking in complex optical potentials},\
  }\href@noop {} {\bibfield  {journal} {\bibinfo  {journal} {Physical Leview
  Letters}\ }\textbf {\bibinfo {volume} {103}},\ \bibinfo {pages} {093902}
  (\bibinfo {year} {2009})}\BibitemShut {NoStop}%
\bibitem [{\citenamefont {Hodaei}\ \emph {et~al.}(2017)\citenamefont {Hodaei},
  \citenamefont {Hassan}, \citenamefont {Wittek}, \citenamefont
  {Garcia-Gracia}, \citenamefont {El-Ganainy}, \citenamefont
  {Christodoulides},\ and\ \citenamefont {Khajavikhan}}]{ar39}%
  \BibitemOpen
  \bibfield  {author} {\bibinfo {author} {\bibfnamefont {H.}~\bibnamefont
  {Hodaei}}, \bibinfo {author} {\bibfnamefont {A.~U.}\ \bibnamefont {Hassan}},
  \bibinfo {author} {\bibfnamefont {S.}~\bibnamefont {Wittek}}, \bibinfo
  {author} {\bibfnamefont {H.}~\bibnamefont {Garcia-Gracia}}, \bibinfo {author}
  {\bibfnamefont {R.}~\bibnamefont {El-Ganainy}}, \bibinfo {author}
  {\bibfnamefont {D.~N.}\ \bibnamefont {Christodoulides}},\ and\ \bibinfo
  {author} {\bibfnamefont {M.}~\bibnamefont {Khajavikhan}},\ }\bibfield
  {title} {\bibinfo {title} {Enhanced sensitivity at higher-order exceptional
  points},\ }\href@noop {} {\bibfield  {journal} {\bibinfo  {journal} {Nature}\
  }\textbf {\bibinfo {volume} {548}},\ \bibinfo {pages} {187} (\bibinfo {year}
  {2017})}\BibitemShut {NoStop}%
\bibitem [{\citenamefont {Bender}\ \emph {et~al.}(2002)\citenamefont {Bender},
  \citenamefont {Brody},\ and\ \citenamefont {Jones}}]{ar14}%
  \BibitemOpen
  \bibfield  {author} {\bibinfo {author} {\bibfnamefont {C.~M.}\ \bibnamefont
  {Bender}}, \bibinfo {author} {\bibfnamefont {D.~C.}\ \bibnamefont {Brody}},\
  and\ \bibinfo {author} {\bibfnamefont {H.~F.}\ \bibnamefont {Jones}},\
  }\bibfield  {title} {\bibinfo {title} {Complex extension of quantum
  mechanics},\ }\href@noop {} {\bibfield  {journal} {\bibinfo  {journal}
  {Physical Review Letters}\ }\textbf {\bibinfo {volume} {89}},\ \bibinfo
  {pages} {270401} (\bibinfo {year} {2002})}\BibitemShut {NoStop}%
\bibitem [{\citenamefont {Bouwmeester}\ \emph {et~al.}(1999)\citenamefont
  {Bouwmeester}, \citenamefont {Pan}, \citenamefont {Daniell}, \citenamefont
  {Weinfurter},\ and\ \citenamefont {Zeilinger}}]{ar38}%
  \BibitemOpen
  \bibfield  {author} {\bibinfo {author} {\bibfnamefont {D.}~\bibnamefont
  {Bouwmeester}}, \bibinfo {author} {\bibfnamefont {J.-W.}\ \bibnamefont
  {Pan}}, \bibinfo {author} {\bibfnamefont {M.}~\bibnamefont {Daniell}},
  \bibinfo {author} {\bibfnamefont {H.}~\bibnamefont {Weinfurter}},\ and\
  \bibinfo {author} {\bibfnamefont {A.}~\bibnamefont {Zeilinger}},\ }\bibfield
  {title} {\bibinfo {title} {Observation of three-photon
  greenberger-horne-zeilinger entanglement},\ }\href@noop {} {\bibfield
  {journal} {\bibinfo  {journal} {Physical Review Letters}\ }\textbf {\bibinfo
  {volume} {82}},\ \bibinfo {pages} {1345} (\bibinfo {year}
  {1999})}\BibitemShut {NoStop}%
\bibitem [{\citenamefont {Gerry}\ and\ \citenamefont {Knight}(2023)}]{ar12}%
  \BibitemOpen
  \bibfield  {author} {\bibinfo {author} {\bibfnamefont {C.~C.}\ \bibnamefont
  {Gerry}}\ and\ \bibinfo {author} {\bibfnamefont {P.~L.}\ \bibnamefont
  {Knight}},\ }\href@noop {} {\emph {\bibinfo {title} {Introductory quantum
  optics}}}\ (\bibinfo  {publisher} {Cambridge university press},\ \bibinfo
  {year} {2023})\BibitemShut {NoStop}%
\bibitem [{\citenamefont {Ku}\ \emph {et~al.}(2018)\citenamefont {Ku},
  \citenamefont {Chen}, \citenamefont {Budroni}, \citenamefont {Miranowicz},
  \citenamefont {Chen},\ and\ \citenamefont {Nori}}]{ar42}%
  \BibitemOpen
  \bibfield  {author} {\bibinfo {author} {\bibfnamefont {H.-Y.}\ \bibnamefont
  {Ku}}, \bibinfo {author} {\bibfnamefont {S.-L.}\ \bibnamefont {Chen}},
  \bibinfo {author} {\bibfnamefont {C.}~\bibnamefont {Budroni}}, \bibinfo
  {author} {\bibfnamefont {A.}~\bibnamefont {Miranowicz}}, \bibinfo {author}
  {\bibfnamefont {Y.-N.}\ \bibnamefont {Chen}},\ and\ \bibinfo {author}
  {\bibfnamefont {F.}~\bibnamefont {Nori}},\ }\bibfield  {title} {\bibinfo
  {title} {Einstein-podolsky-rosen steering: Its geometric quantification and
  witness},\ }\href@noop {} {\bibfield  {journal} {\bibinfo  {journal}
  {Physical Review A}\ }\textbf {\bibinfo {volume} {97}},\ \bibinfo {pages}
  {022338} (\bibinfo {year} {2018})}\BibitemShut {NoStop}%
\bibitem [{\citenamefont {Chitambar}\ and\ \citenamefont {Gour}(2019)}]{ar43}%
  \BibitemOpen
  \bibfield  {author} {\bibinfo {author} {\bibfnamefont {E.}~\bibnamefont
  {Chitambar}}\ and\ \bibinfo {author} {\bibfnamefont {G.}~\bibnamefont
  {Gour}},\ }\bibfield  {title} {\bibinfo {title} {Quantum resource theories},\
  }\href@noop {} {\bibfield  {journal} {\bibinfo  {journal} {Reviews of modern
  physics}\ }\textbf {\bibinfo {volume} {91}},\ \bibinfo {pages} {025001}
  (\bibinfo {year} {2019})}\BibitemShut {NoStop}%
\bibitem [{\citenamefont {Ku}\ \emph {et~al.}(2022)\citenamefont {Ku},
  \citenamefont {Hsieh}, \citenamefont {Chen}, \citenamefont {Chen},\ and\
  \citenamefont {Budroni}}]{ar44}%
  \BibitemOpen
  \bibfield  {author} {\bibinfo {author} {\bibfnamefont {H.-Y.}\ \bibnamefont
  {Ku}}, \bibinfo {author} {\bibfnamefont {C.-Y.}\ \bibnamefont {Hsieh}},
  \bibinfo {author} {\bibfnamefont {S.-L.}\ \bibnamefont {Chen}}, \bibinfo
  {author} {\bibfnamefont {Y.-N.}\ \bibnamefont {Chen}},\ and\ \bibinfo
  {author} {\bibfnamefont {C.}~\bibnamefont {Budroni}},\ }\bibfield  {title}
  {\bibinfo {title} {Complete classification of steerability under local
  filters and its relation with measurement incompatibility},\ }\href@noop {}
  {\bibfield  {journal} {\bibinfo  {journal} {Nature communications}\ }\textbf
  {\bibinfo {volume} {13}},\ \bibinfo {pages} {4973} (\bibinfo {year}
  {2022})}\BibitemShut {NoStop}%
\bibitem [{\citenamefont {Chien}\ \emph {et~al.}(2024)\citenamefont {Chien},
  \citenamefont {Goswami}, \citenamefont {Wu}, \citenamefont {Hiew},
  \citenamefont {Chen},\ and\ \citenamefont {Jen}}]{ar19}%
  \BibitemOpen
  \bibfield  {author} {\bibinfo {author} {\bibfnamefont {C.}~\bibnamefont
  {Chien}}, \bibinfo {author} {\bibfnamefont {S.}~\bibnamefont {Goswami}},
  \bibinfo {author} {\bibfnamefont {C.}~\bibnamefont {Wu}}, \bibinfo {author}
  {\bibfnamefont {W.}~\bibnamefont {Hiew}}, \bibinfo {author} {\bibfnamefont
  {Y.}~\bibnamefont {Chen}},\ and\ \bibinfo {author} {\bibfnamefont {H.~H.}\
  \bibnamefont {Jen}},\ }\bibfield  {title} {\bibinfo {title} {Generating
  scalable graph states in an atom-nanophotonic interface},\ }\href@noop {}
  {\bibfield  {journal} {\bibinfo  {journal} {Quantum Science and Technology}\
  }\textbf {\bibinfo {volume} {9}},\ \bibinfo {pages} {025020} (\bibinfo {year}
  {2024})}\BibitemShut {NoStop}%
\bibitem [{\citenamefont {Zhou}\ \emph {et~al.}(2023)\citenamefont {Zhou},
  \citenamefont {Lu}, \citenamefont {Praquin}, \citenamefont {Chien},
  \citenamefont {Kaufman}, \citenamefont {Cao}, \citenamefont {Xia},
  \citenamefont {Mong}, \citenamefont {Pfaff}, \citenamefont {Pekker} \emph
  {et~al.}}]{ar27}%
  \BibitemOpen
  \bibfield  {author} {\bibinfo {author} {\bibfnamefont {C.}~\bibnamefont
  {Zhou}}, \bibinfo {author} {\bibfnamefont {P.}~\bibnamefont {Lu}}, \bibinfo
  {author} {\bibfnamefont {M.}~\bibnamefont {Praquin}}, \bibinfo {author}
  {\bibfnamefont {T.-C.}\ \bibnamefont {Chien}}, \bibinfo {author}
  {\bibfnamefont {R.}~\bibnamefont {Kaufman}}, \bibinfo {author} {\bibfnamefont
  {X.}~\bibnamefont {Cao}}, \bibinfo {author} {\bibfnamefont {M.}~\bibnamefont
  {Xia}}, \bibinfo {author} {\bibfnamefont {R.~S.}\ \bibnamefont {Mong}},
  \bibinfo {author} {\bibfnamefont {W.}~\bibnamefont {Pfaff}}, \bibinfo
  {author} {\bibfnamefont {D.}~\bibnamefont {Pekker}}, \emph {et~al.},\
  }\bibfield  {title} {\bibinfo {title} {Realizing all-to-all couplings among
  detachable quantum modules using a microwave quantum state router},\
  }\href@noop {} {\bibfield  {journal} {\bibinfo  {journal} {npj Quantum
  Information}\ }\textbf {\bibinfo {volume} {9}},\ \bibinfo {pages} {54}
  (\bibinfo {year} {2023})}\BibitemShut {NoStop}%
\bibitem [{\citenamefont {Roy}\ \emph {et~al.}(2020)\citenamefont {Roy},
  \citenamefont {Hazra}, \citenamefont {Kundu}, \citenamefont {Chand},
  \citenamefont {Patankar},\ and\ \citenamefont {Vijay}}]{ar28}%
  \BibitemOpen
  \bibfield  {author} {\bibinfo {author} {\bibfnamefont {T.}~\bibnamefont
  {Roy}}, \bibinfo {author} {\bibfnamefont {S.}~\bibnamefont {Hazra}}, \bibinfo
  {author} {\bibfnamefont {S.}~\bibnamefont {Kundu}}, \bibinfo {author}
  {\bibfnamefont {M.}~\bibnamefont {Chand}}, \bibinfo {author} {\bibfnamefont
  {M.~P.}\ \bibnamefont {Patankar}},\ and\ \bibinfo {author} {\bibfnamefont
  {R.}~\bibnamefont {Vijay}},\ }\bibfield  {title} {\bibinfo {title}
  {Programmable superconducting processor with native three-qubit gates},\
  }\href@noop {} {\bibfield  {journal} {\bibinfo  {journal} {Physical Review
  Applied}\ }\textbf {\bibinfo {volume} {14}},\ \bibinfo {pages} {014072}
  (\bibinfo {year} {2020})}\BibitemShut {NoStop}%
\bibitem [{\citenamefont {Johansson}\ \emph {et~al.}(2012)\citenamefont
  {Johansson}, \citenamefont {Nation},\ and\ \citenamefont {Nori}}]{ar15}%
  \BibitemOpen
  \bibfield  {author} {\bibinfo {author} {\bibfnamefont {J.~R.}\ \bibnamefont
  {Johansson}}, \bibinfo {author} {\bibfnamefont {P.~D.}\ \bibnamefont
  {Nation}},\ and\ \bibinfo {author} {\bibfnamefont {F.}~\bibnamefont {Nori}},\
  }\bibfield  {title} {\bibinfo {title} {Qutip: An open-source python framework
  for the dynamics of open quantum systems},\ }\href@noop {} {\bibfield
  {journal} {\bibinfo  {journal} {Computer Physics Communications}\ }\textbf
  {\bibinfo {volume} {183}},\ \bibinfo {pages} {1760} (\bibinfo {year}
  {2012})}\BibitemShut {NoStop}%
\bibitem [{\citenamefont {Wootters}(1998)}]{ar20}%
  \BibitemOpen
  \bibfield  {author} {\bibinfo {author} {\bibfnamefont {W.~K.}\ \bibnamefont
  {Wootters}},\ }\bibfield  {title} {\bibinfo {title} {Entanglement of
  formation of an arbitrary state of two qubits},\ }\href@noop {} {\bibfield
  {journal} {\bibinfo  {journal} {Physical Review Letters}\ }\textbf {\bibinfo
  {volume} {80}},\ \bibinfo {pages} {2245} (\bibinfo {year}
  {1998})}\BibitemShut {NoStop}%
\bibitem [{\citenamefont {Ge}\ \emph {et~al.}(2023)\citenamefont {Ge},
  \citenamefont {Liu},\ and\ \citenamefont {Cheng}}]{ar9}%
  \BibitemOpen
  \bibfield  {author} {\bibinfo {author} {\bibfnamefont {X.}~\bibnamefont
  {Ge}}, \bibinfo {author} {\bibfnamefont {L.}~\bibnamefont {Liu}},\ and\
  \bibinfo {author} {\bibfnamefont {S.}~\bibnamefont {Cheng}},\ }\bibfield
  {title} {\bibinfo {title} {Tripartite entanglement measure under local
  operations and classical communication},\ }\href@noop {} {\bibfield
  {journal} {\bibinfo  {journal} {Physical Review A}\ }\textbf {\bibinfo
  {volume} {107}},\ \bibinfo {pages} {032405} (\bibinfo {year}
  {2023})}\BibitemShut {NoStop}%
\end{thebibliography}%
\begin{acknowledgments}
H. H. J. acknowledges support from the National Science and Technology Council (NSTC), Taiwan, under the Grant No. NSTC-112-2119-M-001-007, and from Academia Sinica under Grant AS-CDA-113-M04. We are also grateful for support from TG 1.2 of NCTS at NTU. H.- Y. K. is supported by the Ministry of Science and Technology, Taiwan, (with grant number MOST 112-2112-M-003- 020-MY3), and Higher Education Sprout Project of National Taiwan Normal University (NTNU).
\end{acknowledgments}
\section*{Author contributions}
H. Y. K. initiated the ideas, J. S. Y., H. Y. K., and H. H. J. supervised the project. C. G. F conducted the analytical and numerical calculations, interpreted the results, and wrote the first draft of the manunscript. All authors contributed to the writing of the manuscript.
\section*{Competing Interests}
The authors declare no competing interests.

\end{document}